\documentclass[a4paper]{article}
\usepackage{graphicx}

\pdfoutput=1

\begin{document}

\begin{center}
{\Large \bf Toward Description of $pp$ and $p{\rm C}$ Interactions at High Energies:
            Problems of Fritiof-based Models}
\end{center}

\begin{center}
{V. Uzhinsky}
\end{center}

\begin{center}
{Laboratory of Information Technologies, JINR, Dubna, Russia}
\end{center}

\begin{center}
\begin{minipage}{14cm}
How do the models (Fritiof 1.6, Fritiof 7.0, UrQMD 3.3 and Hijing 1.383) describe
experimental data of NA61/SHINE and NA49 Collaborations on $pp$ and $p{\rm C}$
interactions at high energies? An answer on this question is given in the paper.
It is shown that the UrQMD model does not reproduce the energy dependence of
$\pi^-$ meson production in $pp$-interactions; the Fritiof 1.6 and the Fritiof 7.0
models underestimate the meson production on $\sim$ 10 \%; the Hijing model
overestimates the data on $\sim$ 10 \%.
A change of a LUND fragmentation function parameter in the Fritiof models 1.6/7.0
allows to describe the data. A decreasing of probabilities of binary processes
in the UrQMD model like $p+p\rightarrow N+N'$ and so on, allows one to
describe the data, thus a problem of a correct accounting of the processes
arises. An increasing of a probability of the single diffraction dissociation
in the Hijing model from 35 \% to 50 \% allows to describe the data, though
a spectrum of masses produced in the diffraction is not satisfactory. A description
of the diffraction is a problem in all the models.
\end{minipage}
\end{center}

\section*{Introduction}
The NA61/SHINE Collaboration has presented a high precision data on $\pi^-$-meson
inclusive distributions in $pp$ interactions at $P_{lab}=$ 20, 31, 40, 80 and 158 GeV/c
\cite{NA61}. The NA49 Collaboration has published data on $\pi^\pm$, $K^\pm$, proton
and antiproton spectra in $pp$ \cite{NA49pp2pi,NA49pp2p,NA49pp2K}
and $p{\rm C}$ \cite{NA49pC} interactions at 158 GeV/c. The NA61/SHINE Collaboration
has the analogous data for $p{\rm C}$ interactions at 31 GeV/c \cite{NA61pC}.
There were some attempts to describe the last data in the modern Monte Carlo models --
FLUKA, VENUS, UrQMD (see \cite{NA61pC}) and FTF of Geant4 \cite{UzhiImp,UzhiTune}.
Recently, a description of the $pp$ data \cite{NA61,NA49pp2pi,NA49pp2p,NA49pp2K}
in the UrQMD model was given in paper \cite{UzhiUrQ}. As it was shown there,
the model does not reproduce energy dependence of the $\pi^-$-meson production
in the central region (see Fig.~1). A natural question arises --
How do the other models describe the data? In the presented paper, Fritiof-based
models will be considered.
\begin{center}
\begin{figure}[cbth]
\includegraphics[width=75mm,height=45mm,clip]{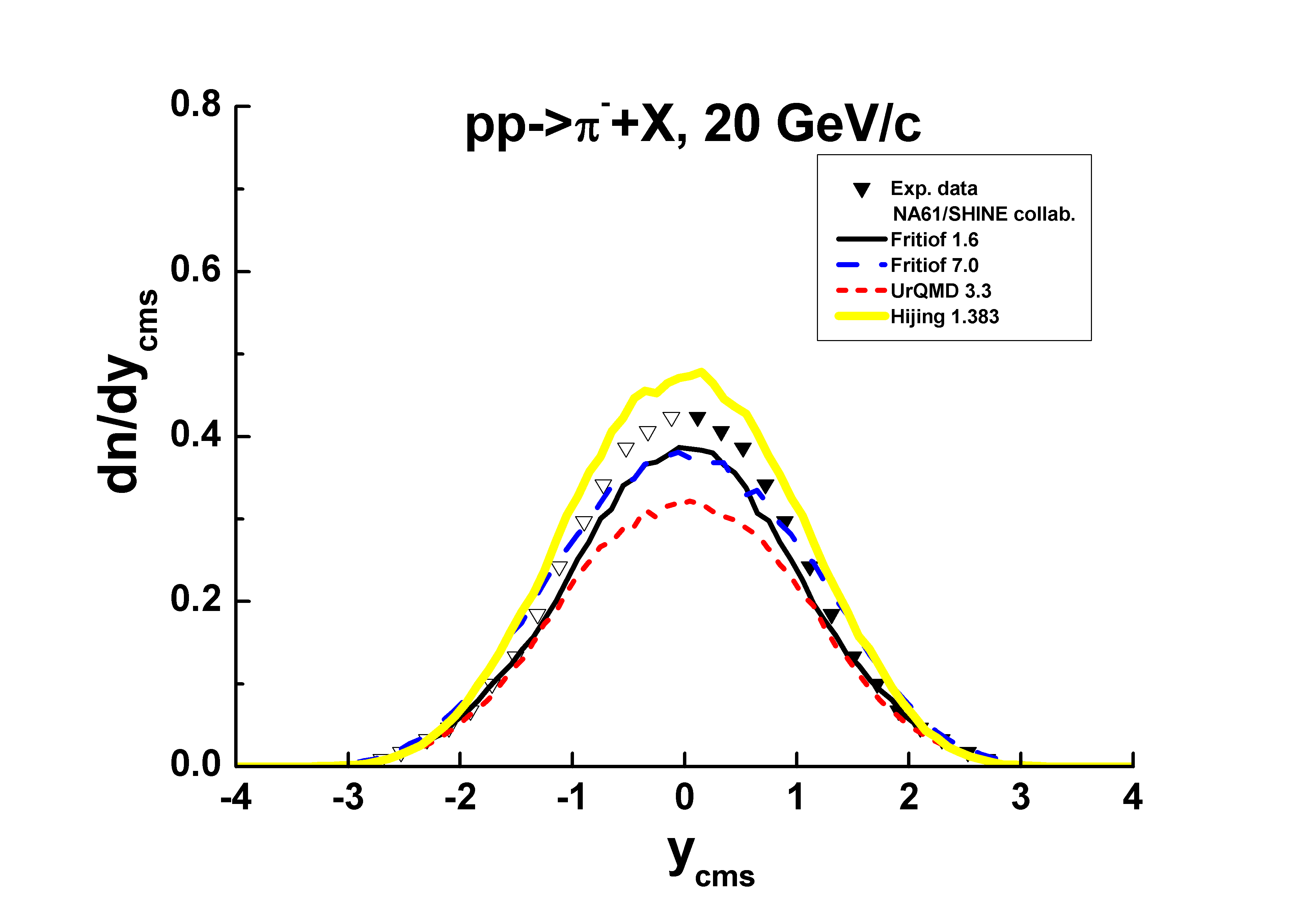}\includegraphics[width=75mm,height=45mm,clip]{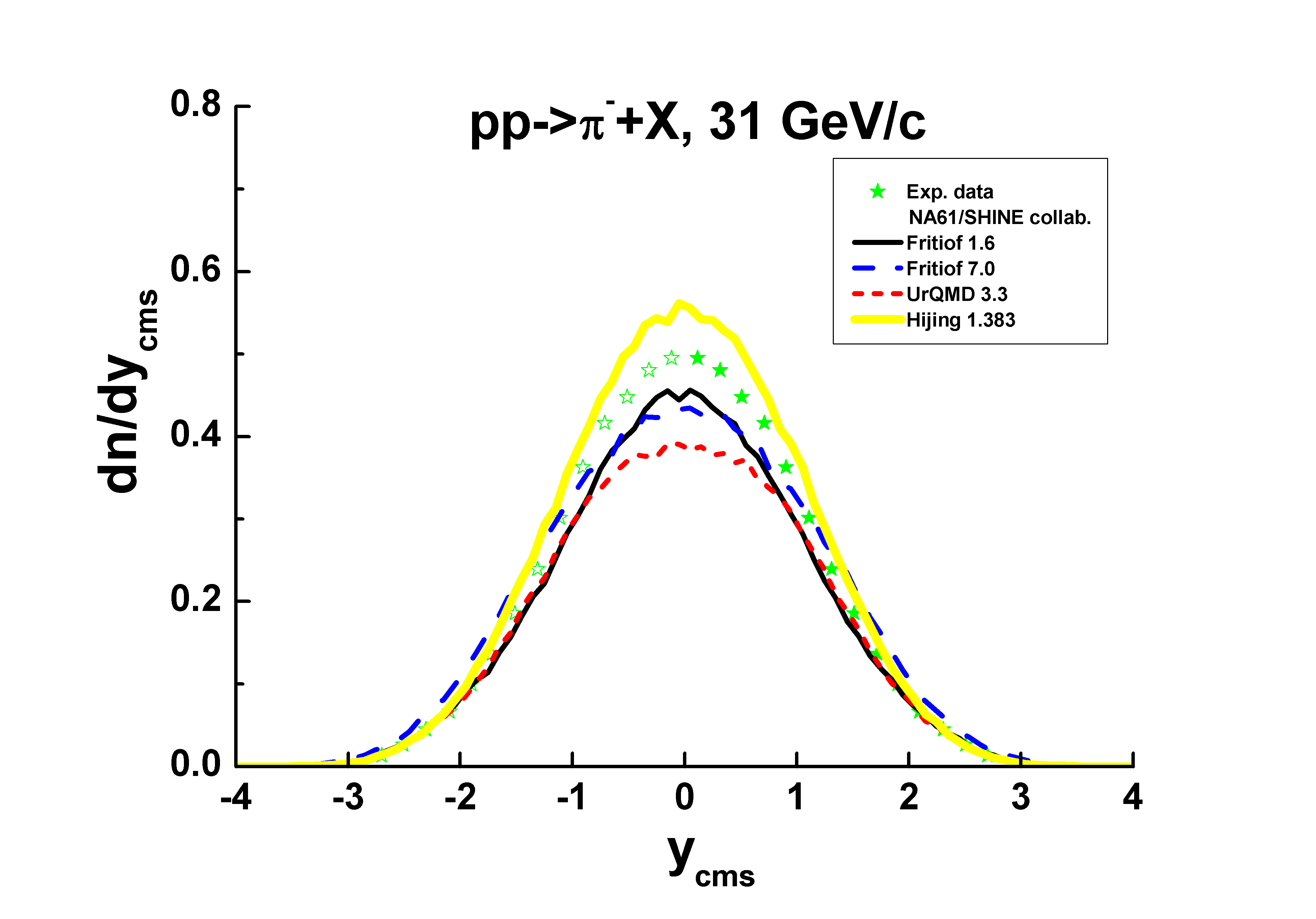}
\includegraphics[width=75mm,height=45mm,clip]{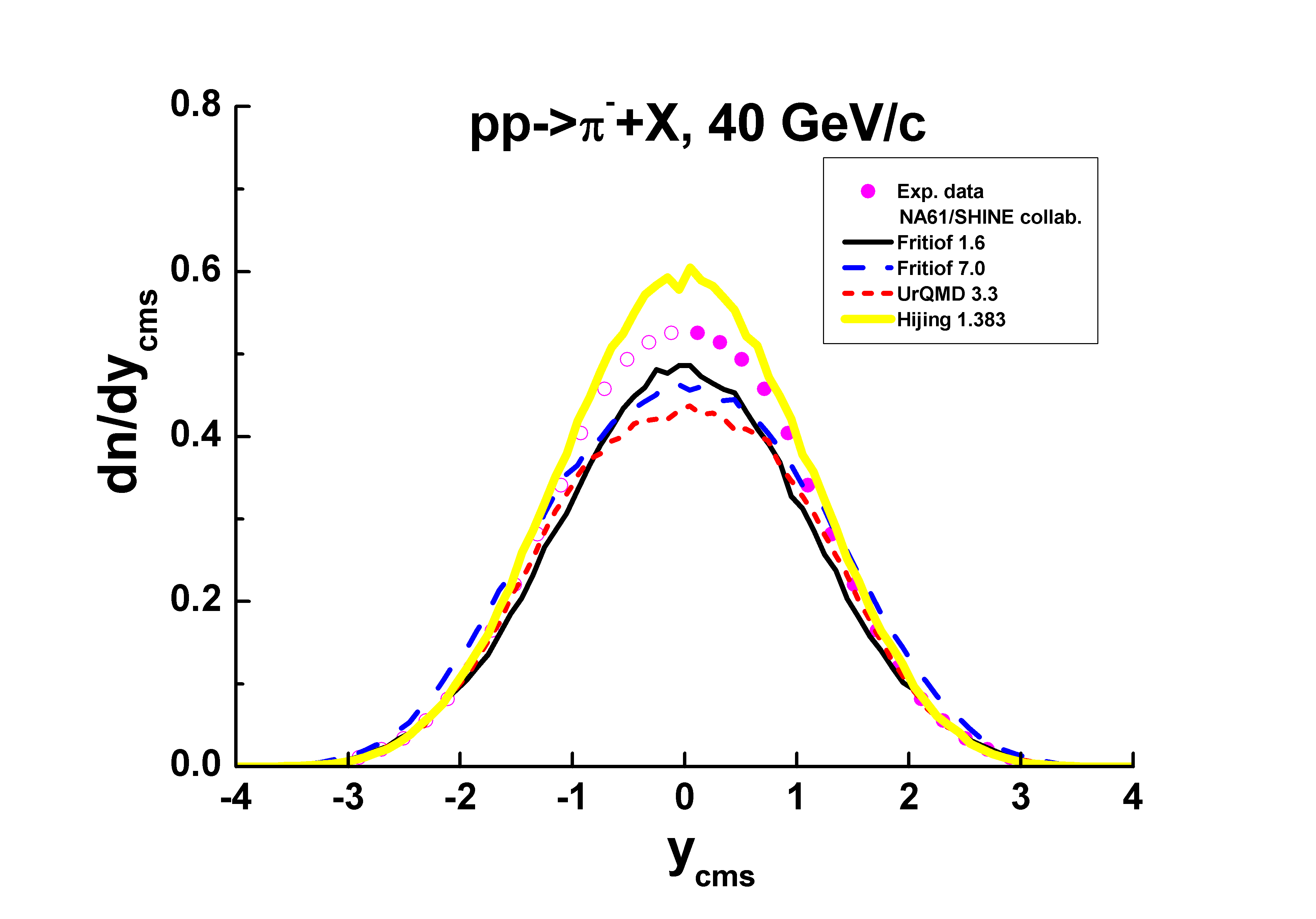}\includegraphics[width=75mm,height=45mm,clip]{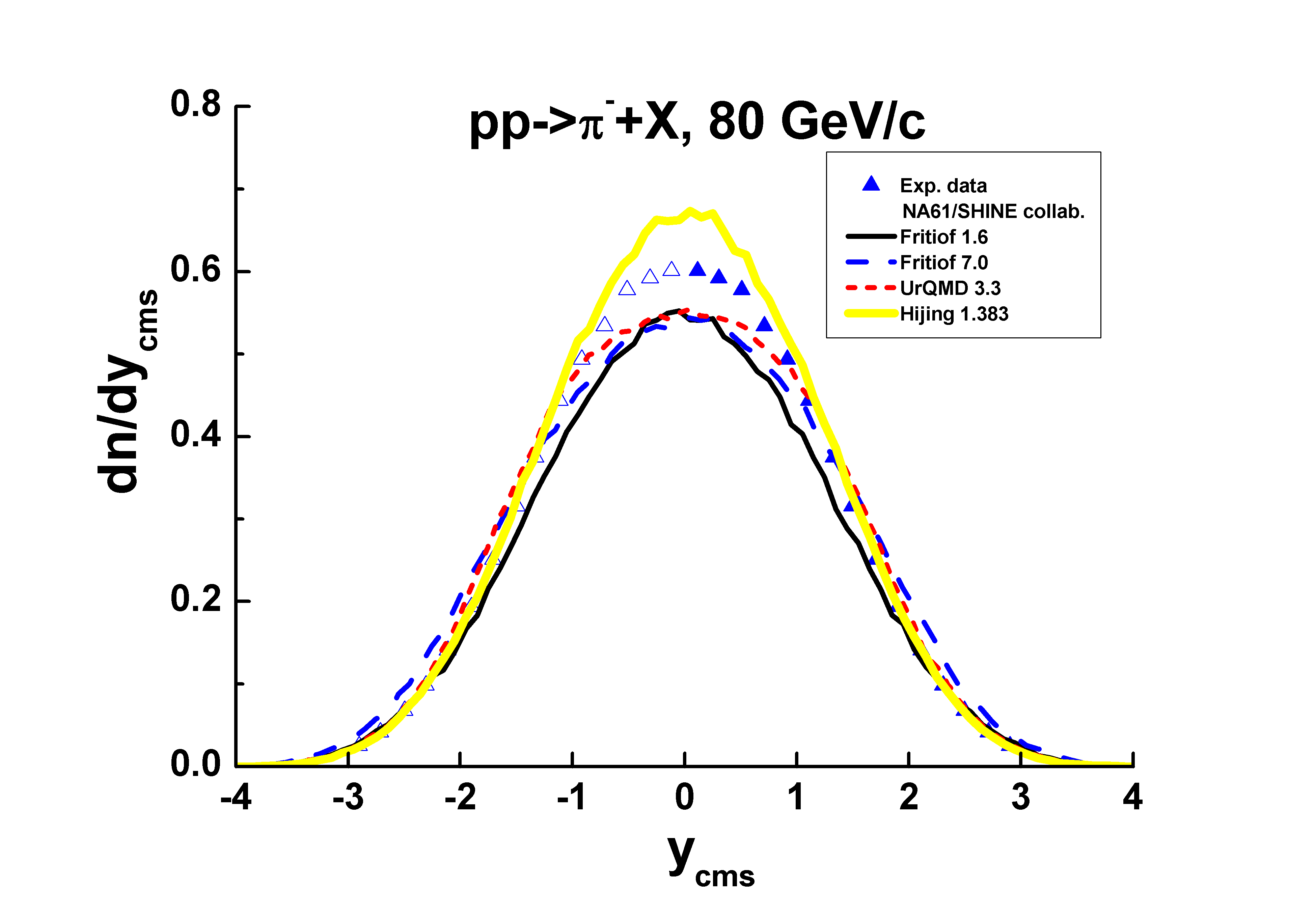}
\caption{Rapidity distributions of $\pi^-$-mesons in $pp$ interactions. Closed points
are the NA61/SHINE experimental data \protect{\cite{NA61}}, the open points are the data
reflected at mid-rapidity. Lines are model calculations.}
\label{NA61pim}
\end{figure}
\end{center}
\vspace{-10mm}
~~~~~~
The Fritiof model \cite{Fritiof1,Fritiof2} is in the basis of well-known event
generators, Hijing and UrQMD. It is explained by its clear physical ideas, and
a defined beauty of its code \cite{Fritiof2} by itself. The Hijing model
\cite{Hijing1,Hijing2} was used at design of RHIC and LHC setups for a study of
nucleus-nucleus interactions at super high energies. Validation of the model is
presented at a web-page \cite{HEPWEBh,HEPWEB}. The Ultra-relativistic Quantum
Molecular Dynamic model (UrQMD) \cite{UrQMD}\footnote{$^)$
The code of the model version 3.3 see at \protect{\cite{UrQMDpage}}.}$^,$
\footnote{$^)$ Validation of the model see at a web-page \protect{\cite{HEPWEBu}}.}$^)$
is now applied at a design of future experiments at FAIR \cite{FAIR} and NICA
\cite{NICA} facilities. The Fritiof model is also implemented in the Geant4 toolkit
\cite{Geant4} under the name -- FTF model. Thus, correctness of the Fritiof-based
models is very important for experimental studies and practical applications.

As it is shown in Fig.~\ref{NA61pim} the models give various predictions for $\pi^-$-meson
rapidity spectra in $pp$ interactions. As seen, the Hijing model overestimates the spectra
in the central region. The UrQMD model does not reproduce the energy dependence of
the spectra. Only at $P_{lab}=$158 GeV/c there is an agreement between the UrQMD model
calculations and the NA49 data. The Fritiof 1.6 and 7.0 models underestimate the data
on 10 \%. In all calculations, the yield of $K^0_s$ decays into $\pi$-meson production
($\sim 10$\%) was not accounted.

\begin{center}
\begin{figure}[cbth]
\includegraphics[width=160mm,height=100mm,clip]{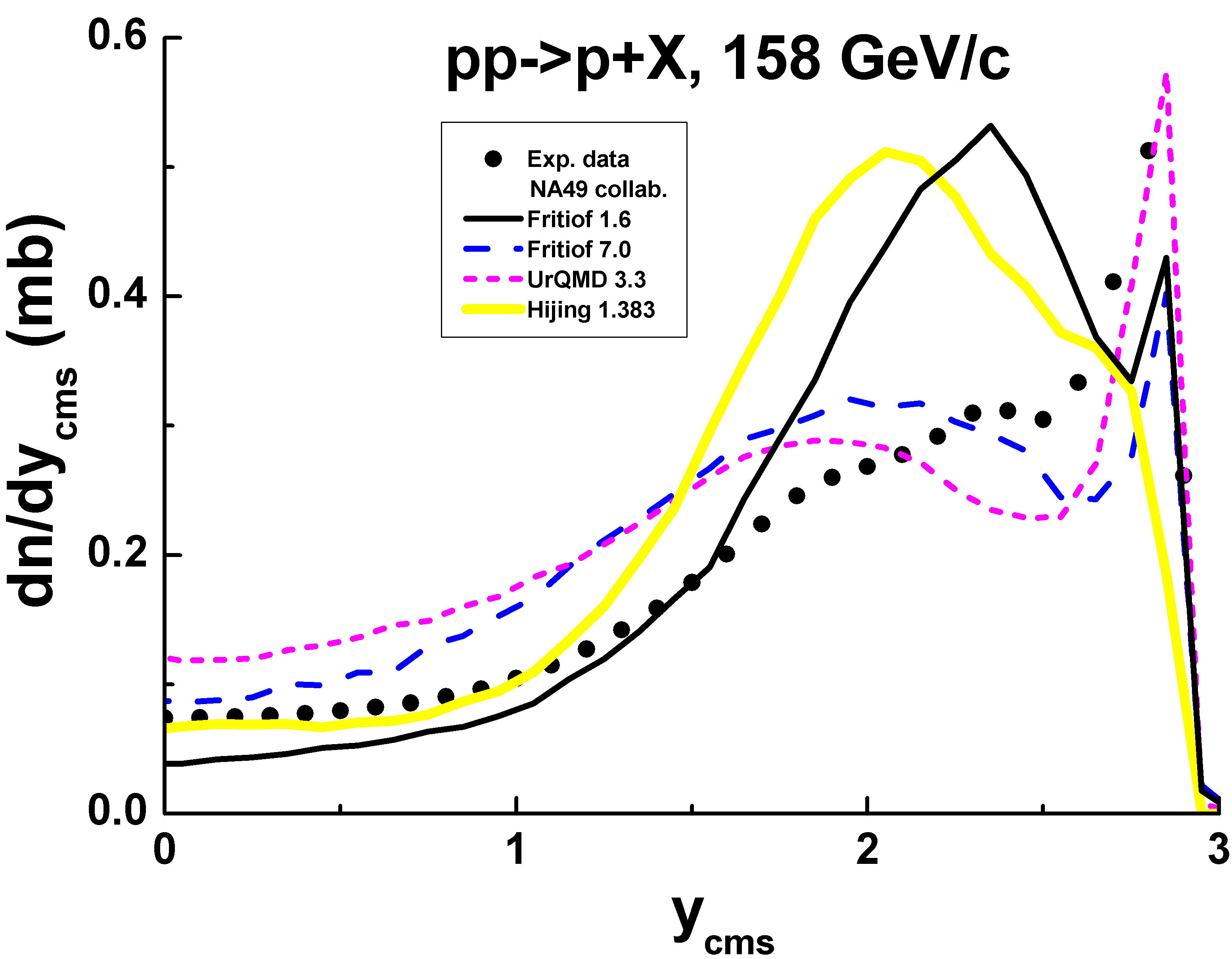}
\caption{Rapidity distributions of protons in $pp$ interactions at $P_{lab}=$ 158 GeV/c.
Points are experimental data \protect{\cite{NA49pp2p}}. Lines are model calculations.}
\label{pp158prot}
\end{figure}
\end{center}

The worst situation takes place with a description of proton spectra presented in
Fig.~\ref{pp158prot}. As seen, the models essentially underestimate a height of
the diffraction peak at $y_{cms}\sim 2.8$ except the UrQMD model. Outside the peak,
the models give various predictions.
\vspace{-10mm}
~~~~~~
In order to understand such results one needs to know the main ideas of the Fritiof
model and their program implementations. The main ideas will be presented in Sec.~1,
and the program implementations are considered in Sec.~2.

Possibilities of model's improvements are considered in Sec.~3. All of them include
a change of a single diffraction dissociation cross section. Necessity of the change
is studied in Sec.~4 where experimental data on the diffraction dissociation and
model's calculations are analyzed. Results for $p{\rm C}$ interactions are given in
Sec.~5.

A general conclusion is -- simulations of the diffraction dissociation and their
cross sections must be improved in the model.

\section{Main ideas of the Fritiof model}
\begin{figure}[cbth]
    \begin{center}
        \includegraphics[width=80mm,height=30mm,clip]{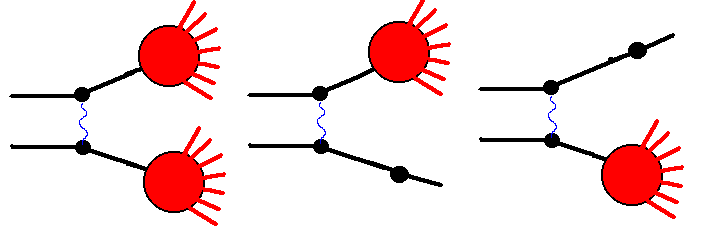}
        \caption{"Non-diffractive" and diffractive interactions considered in the Fritiof model.}
        \label{FTFproc}
    \end{center}
\end{figure}

The Fritiof model \cite{Fritiof1,Fritiof2} assumes that all hadron-hadron
interactions are binary reactions, $h_1+h_2\rightarrow h_1'+h_2'$, where $h_1'$ and
$h_2'$ are excited states of the hadrons with discrete or continuous mass spectra
(see Fig.~\ref{FTFproc}). If one of the final hadrons is in its ground state
($h_1+h_2\rightarrow h_1+h_2'$) the reaction is called "single diffraction dissociation",
and if neither hadron is in its ground state it is called a "non-diffractive" interaction.
\vspace{-5mm}

The excited hadrons are treated as QCD-strings, and the corresponding LUND-string
fragmentation model is applied in order to simulate their decays.

\begin{figure}[cbth]
    \begin{center}
        \includegraphics[width=130mm,height=90mm,clip]{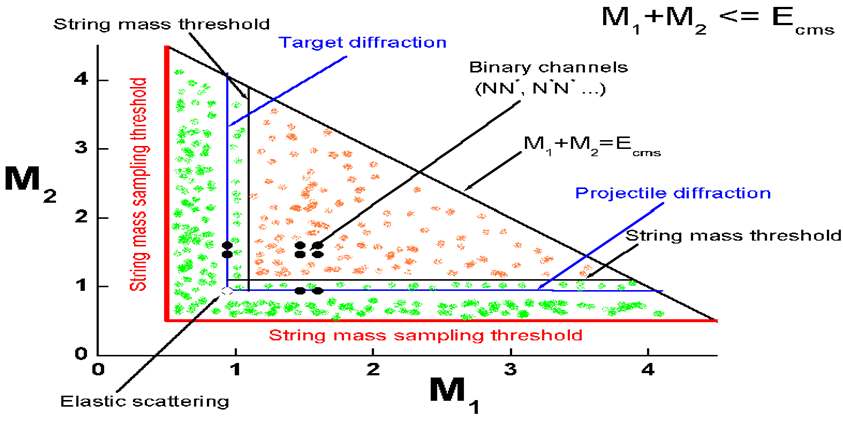}
        \caption{Diagram of the final states of hadron-hadron interactions.}
        \label{Diag}
    \end{center}
\end{figure}

The key ingredient of the Fritiof model is a sampling of the string masses. In general,
the set of final state of interactions can be represented by Fig.~\ref{Diag}, where
samples of possible string masses are shown. There is a point corresponding to elastic
scattering, a group of points which represents final states of binary hadron-hadron
interactions like $N+N\rightarrow N+N^*(1440)$ and so on, lines corresponding to
the diffractive interactions, and various intermediate regions. The region populated
with the red points is responsible for the non-diffractive interactions. In the model,
the mass sampling threshold is set equal to the ground state hadron masses, but in
principle the threshold can be lower than these masses.  The string masses are sampled
in the triangular region restricted by the diagonal line corresponding
to the kinematical limit $M_1+M_2=E_{cms}$ where $M_1$ and $M_2$ are the masses of
the $h_1'$ and $h_2'$ hadrons, and also of the threshold lines. If a point is below
the string mass threshold, $M_d$, it is shifted to the nearest diffraction line.
\vspace{-5mm}

The original Fritiof model had no points corresponding to elastic scattering
and to the binary final states. As it was known at the time of its creation,
the mass of an object produced by diffraction dissociation, $M_x$, for example
from the reaction $p+p\rightarrow p+X$, is distributed as $dM_x/M_x\propto dM^2_x/M^2_x$,
so it was natural to assume that the object mass distributions in all inelastic
interactions obeyed the same law. This can be re-written using the light-cone
momentum variables, $P^+$ or $P^-$, $$P^+=E+p_z,\ \ \ P^-=E-p_z,$$
where $E$ is an energy of a particle, and $p_z$ is its longitudinal momentum along
the collision axis. At large energy and positive $p_z$, $P^-\simeq (M^2+P_T^2)/2p_z$.
At negative $p_z$, $P^+\simeq (M^2+P_T^2)/2|p_z|$. Usually, the transferred transverse
momentum, $P_T$, is small and can be neglected. In the case, $P^-\propto M^2$
and $P^+\propto M^2$. Thus, it was assumed that $P^-$ and $P^+$
of a projectile and a target associated hadron, respectively, are distributed as
\begin{equation}dP^-/P^-,\ \ \ dP^+/P^+.\label{FTFmass}\end{equation}
A gaussian distribution was used for a sampling of $P_T$.

In the case of hadron-nucleus or nucleus-nucleus interactions it was assumed that
the created objects can interact further with other nuclear nucleons and create new
objects. Assuming equal masses of the objects, the multiplicity of particles produced in
these interactions will be proportional to the number of participating nuclear nucleons.
Due to this, the multiplicity of particles produced in hadron-nucleus or nucleus-nucleus
interactions is larger than that in hadron-hadron ones. The probabilities of multiple
intra-nuclear collisions were sampled with the help of a simplified Glauber model.
Cascading of secondary particles was not considered.

Because the Fermi motion of nuclear nucleons was simulated in a simple manner,
the original Fritiof model \cite{Fritiof2} could not work at $P_{lab} <$ 10--20 GeV/c.

It was assumed in the model that the created objects are quark-gluon strings with
constituent quarks at their ends originating from the primary colliding hadrons.
The LUND string fragmentation model (JETSET 6.3 \cite{JETSET6_2}) was applied for
a simulation  of the object decays. It was assumed also that the strings with
sufficiently large masses have "kinks" -- additional radiated gluons (see subroutine
TORSTE in the program code \cite{Fritiof2}). This was very important for a correct
reproduction of particle multiplicities in the interactions.

\section{Program implementations of the Fritiof model}
\subsection*{Fritiof 1.6}
In this model implementation $M_d=$ 1.2 GeV for nucleons and the average
$P_T=$ 0.282 GeV/c. All objects are considered as strings and processed by
JETSET 6.3 {\bf independently}!

It is rather easy to improve the description of the data of Fig.~1 changing
the parameter "a" in the LUND string fragmentation function,
$ f(z)\propto (1/z)\ (1-z)^a\ exp(-bm_T^2/z)$,
from 0.5 to 1.25 (PAR(31)=1.25).

Rather complicated algorithm of a baryon production by a $qq-q$ string
is used in the LUND model. It is implemented in JETSET 6.3 and JETSET 7.3.

\subsection*{Fritiof 7.0}
$M_d=$ 1.2 GeV and the average $P_T=$ 0.1 GeV/c in this model implementation.
Due to the lower $P_T$ than in the original model, the diffraction peak
(see Fig.~\ref{pp158prot}) is more narrow than in the Fritiof 1.6 model.
Strings are not considered, instead quarks of the objects, and sometimes saved
hadrons and gluons are processed by the JETSET 7.3 \cite{JETSET7_3} as a unit
system. Thus, there can be in the system after a diffraction dissociation
a saved baryon, a quark and a diquark. Rather often, there can be a gluon.
If the system mass is small enough ($\leq$ 3 GeV), JETSET 7.3 erases the gluon,
and projects the quark and the diquark on a nearest baryonic state. Very often
the state is $\Delta^+(1232)$. After a sampling of the $\Delta$ mass, a momentum
of the saved baryon is re-defined. This is reflecting on spectra of saved hadrons.

For an inclusion of hard interaction effects and gluonic radiations, the Fritiof model
7.0 is coupled with Pythia 5.5 \cite{Pythia} and Ariadne \cite{Ariadne} codes. Another
inclusion is presented in the Hijing model \cite{Hijing2}.

\subsection*{UrQMD 3.3}
In the code of the model, $M_d=m_N+\delta$ where $m_N$ is a nucleon mass, and
$\delta=0.52$ GeV (CTParam(2)=0.52). The average $P_T=$ 1.6 GeV/c (CTParam(31)=1.6).
Such large value of $P_T$ is caused by another algorithm of the mass
sampling (see subroutine STREXCT in make22.f of the corresponding source file).

Instead of the complicated algorithm of a baryon production in the LUND model implemented
in JETSET 6.3 and JETSET 7.3, the following fragmentation function of a $qq-q$ string
into a baryon is used:
\begin{equation}
f(z)\propto exp\left[-\frac{(z-b)^2}{2a^2}\right],~~~~ a=0.275,~~~ b=0.42,
\label{UrQMDff}
\end{equation}
%\begin{equation}
%f(z)\propto (z-z_{min})^2/(z_{max}-z_{min})^2,
%\end{equation}
%\begin{equation}
%f(z)\propto 1/\sqrt{z}
%\end{equation}
where $z$ is a fraction of light-code momentum of the string given to a created baryon.

There is a possibility to consider a diquark as a unit in other Fritiof model
implementations setting, for example, an option MST(10)=0 in JETSET 6.3. In the case,
the Feynman-Field fragmentation function will be used for diquarks.

UrQMD models considers also the following binary reaction:
$N+N\rightarrow N+\Delta(1232)$,
$N+N\rightarrow N+N^*$,
$N+N\rightarrow N+\Delta^*$,
$N+N\rightarrow \Delta(1232)+\Delta(1232)$,
$N+N\rightarrow \Delta(1232)+N^*$,
$N+N\rightarrow \Delta(1232)+\Delta^*,~~~\Delta(1232)+N^*$,
$N+N\rightarrow N^*+N^*,~~~\Delta^*+N^*,~~\Delta(1232)+N^*,~~~\Delta^*+\Delta^*$.
They are very important at low energies.

\subsection*{Hijing 1.383}
In the model, $M_d=m_{qq}+\delta$ where $m_{qq}\sim 770$ MeV is a di-quark mass, and
$\delta=1.5$ GeV (HIPR1(1)=1.5). Though, the value is not used for a separation of
the diffractive and non-diffractive interactions. Instead if this, a probability of
the single diffraction dissociation for nucleon-nucleon interactions is setting to
35~\% for the considered energies (see BLOCK DATA HIDATA, line --
DATA (HIDAT0(4,I),I=1,10)/0.35, 0.35, 0.3, 0.3, 0.3, 0.3,), and a mass of a diffractive
produced system is sampled along the diffraction lines of Fig.~3 (see SUBROUTINE HIJSFT,
lines -- 222~~~~~~X2=HIRND2(6,XMIN,XMAX), and 242~~~~~~X1=HIRND2(6,XMIN,XMAX)). A lower
bound of the mass is about 2 GeV.

A creation of kinky strings is implemented in the model as well as a production of jets
and mini-jets. But they are not important for our aims.

\section{Tuning of the models}
\subsection{Hijing model with increased probability of the diffraction dissociation}
%\begin{center}
\begin{figure}[cbth]
\includegraphics[width=75mm,height=25mm,clip]{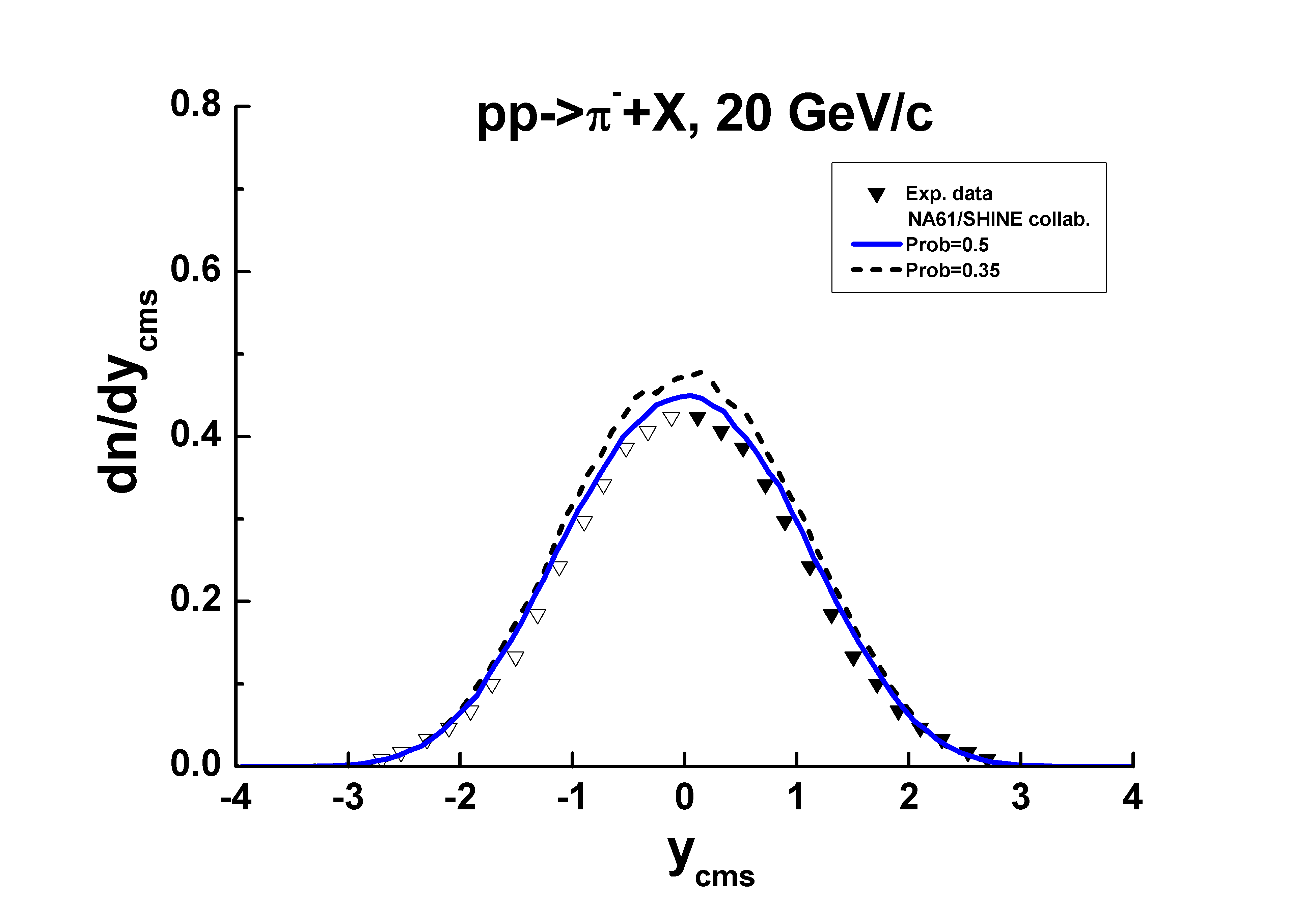}\includegraphics[width=75mm,height=25mm,clip]{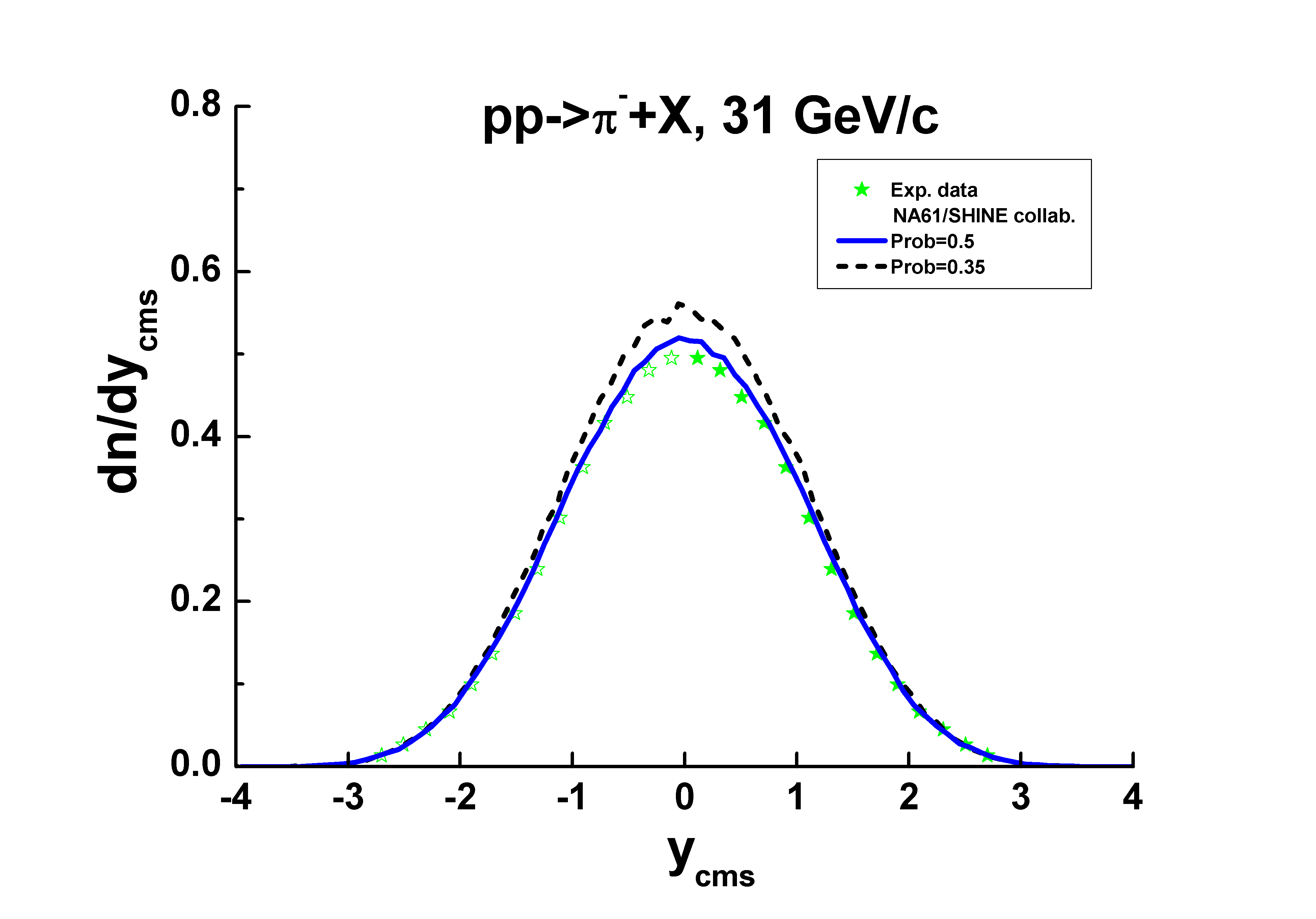}
\includegraphics[width=75mm,height=25mm,clip]{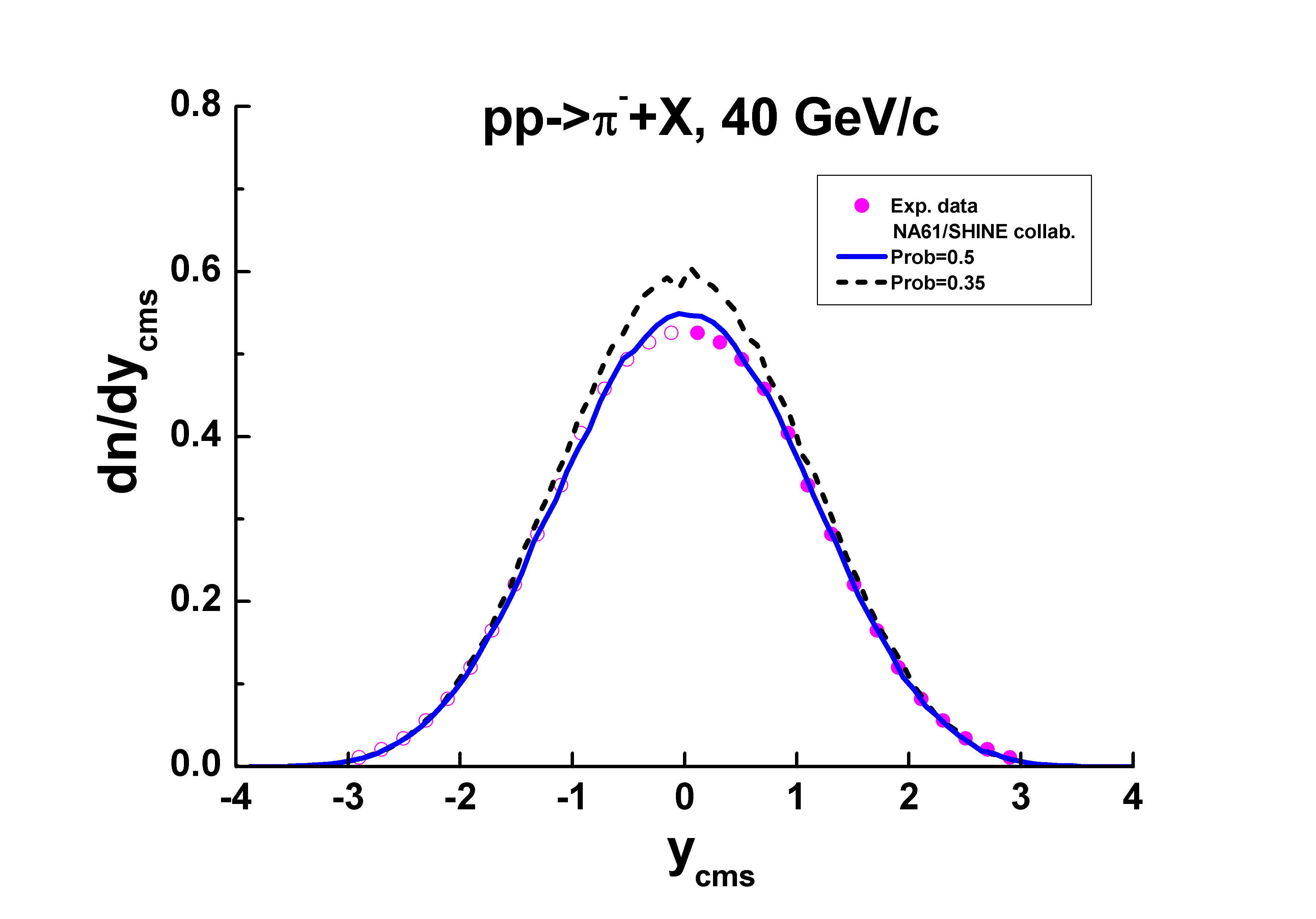}\includegraphics[width=75mm,height=25mm,clip]{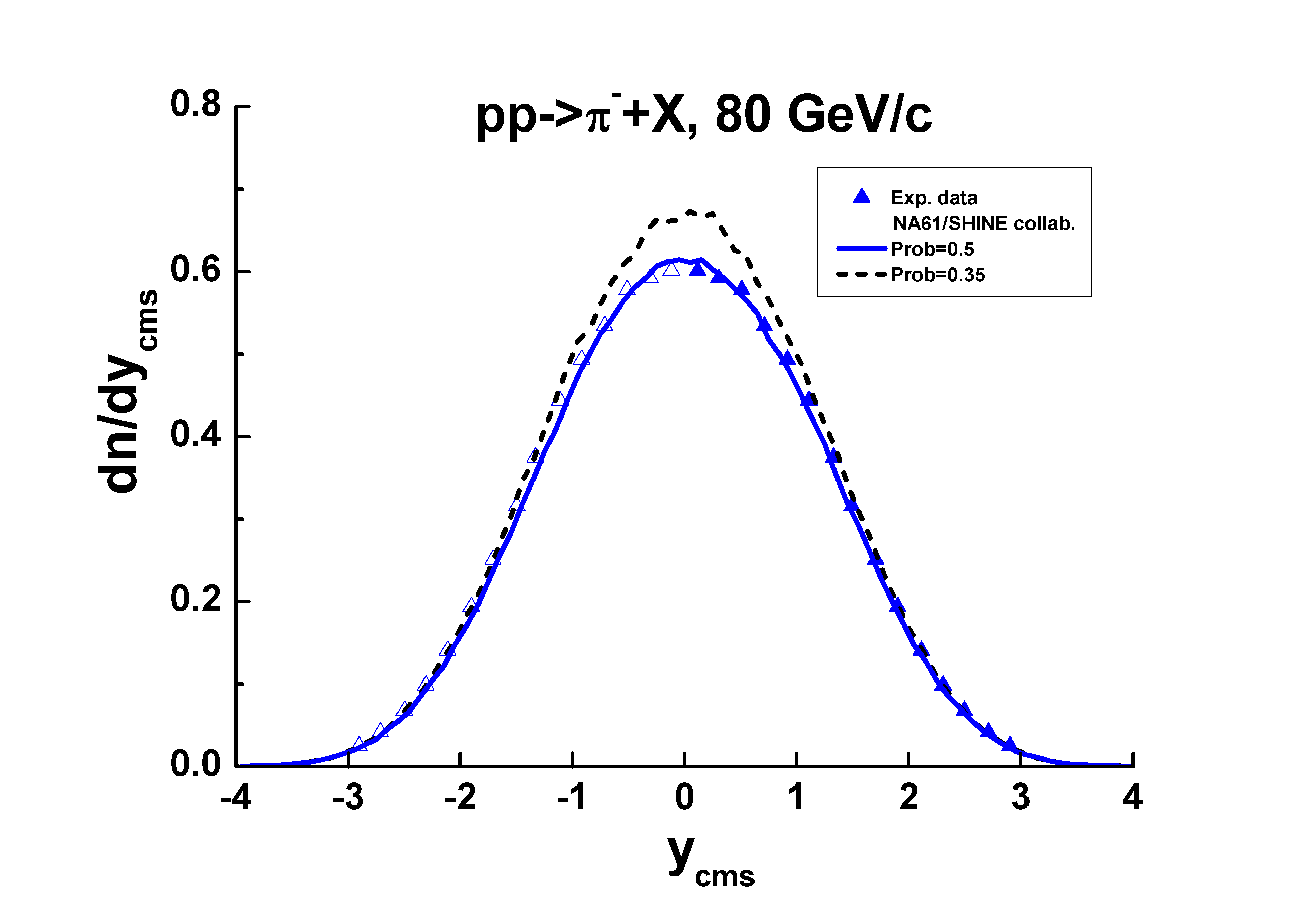}
\caption{Rapidity distributions of $\pi^-$-mesons in $pp$ interactions. Closed points
are the NA61/SHINE experimental data \protect{\cite{NA61}}, the open points are the data
reflected at mid-rapidity. Lines are the Hijing model calculations with and without increasing
of the probability of the diffraction (solid and dashed lines, respectively).}
\label{NA61hij}
\end{figure}
%\end{center}
\vspace{-3mm}
A simple way to decrease the Hijing model results and reach an agreement with
the experimental data is an increasing of the probability of the diffraction
dissociation. This can be done inserting a change in BLOCK DATA HIDATA:
\begin{verbatim}
*   DATA (HIDAT0(4,I),I=1,10)/0.35,0.35,0.3,0.3,0.3,0.3,   ! Uzhi
    DATA (HIDAT0(4,I),I=1,10)/0.5 ,0.5 ,0.3,0.3,0.3,0.3,   ! Uzhi
\end{verbatim}

The results are presented in Fig.~\ref{NA61hij}. As seen, there is a good opportunity to improve
the model results tuning more exactly the probability.

\subsection{Tuning of the Fritiof 1.6 model}
\begin{figure}[cbth]
\includegraphics[width=50mm,height=55mm,clip]{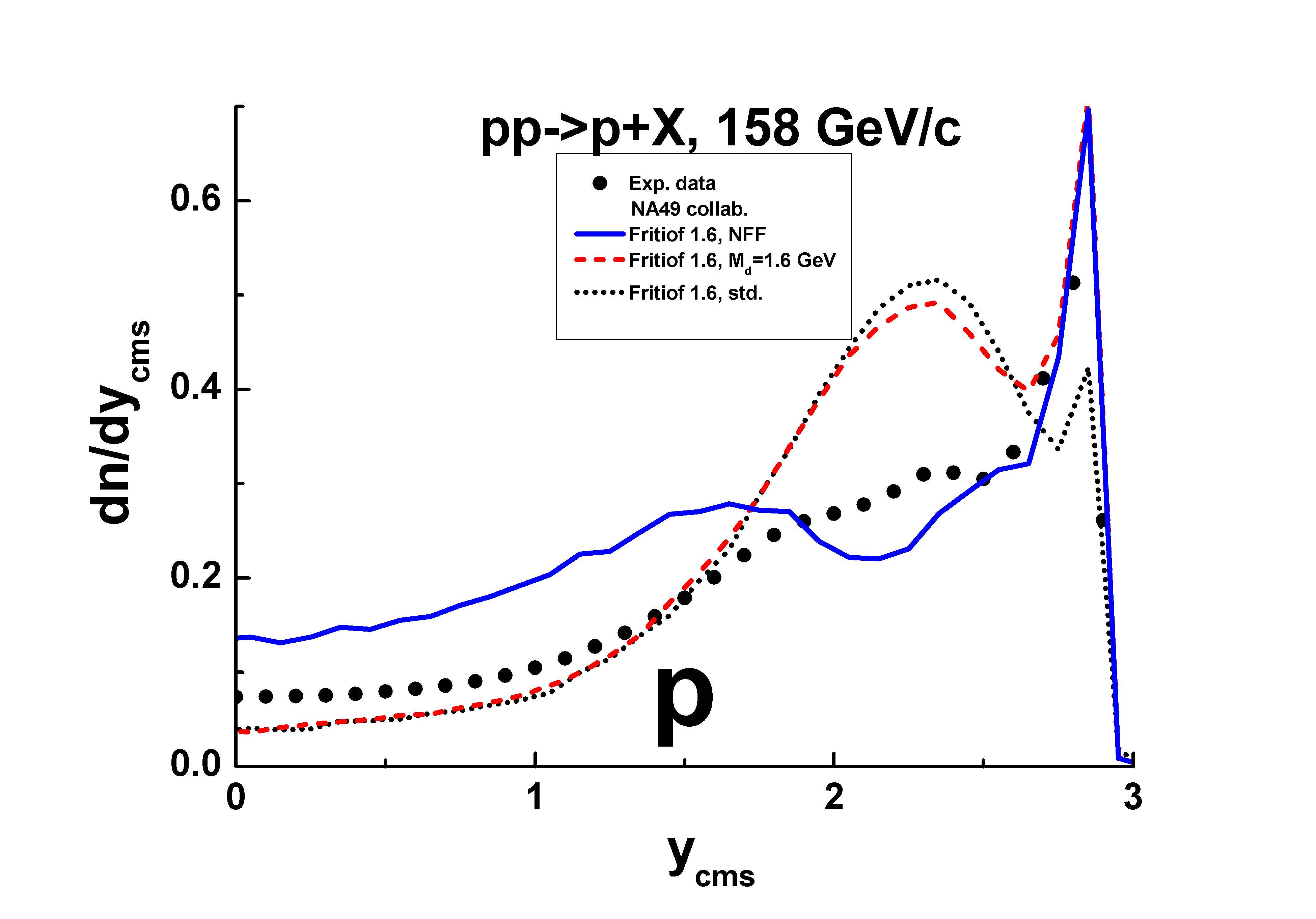}\includegraphics[width=50mm,height=55mm,clip]{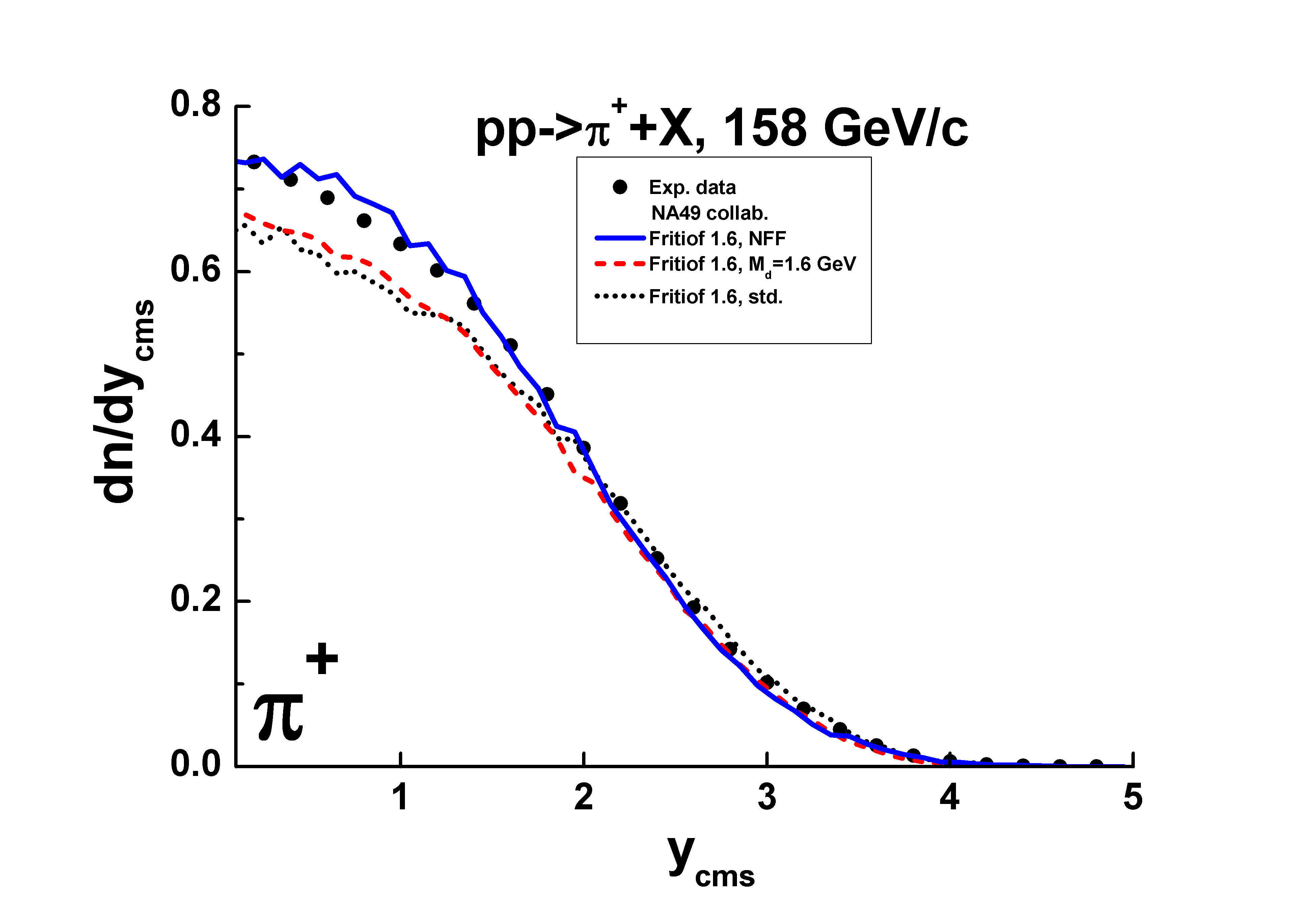}\includegraphics[width=50mm,height=55mm,clip]{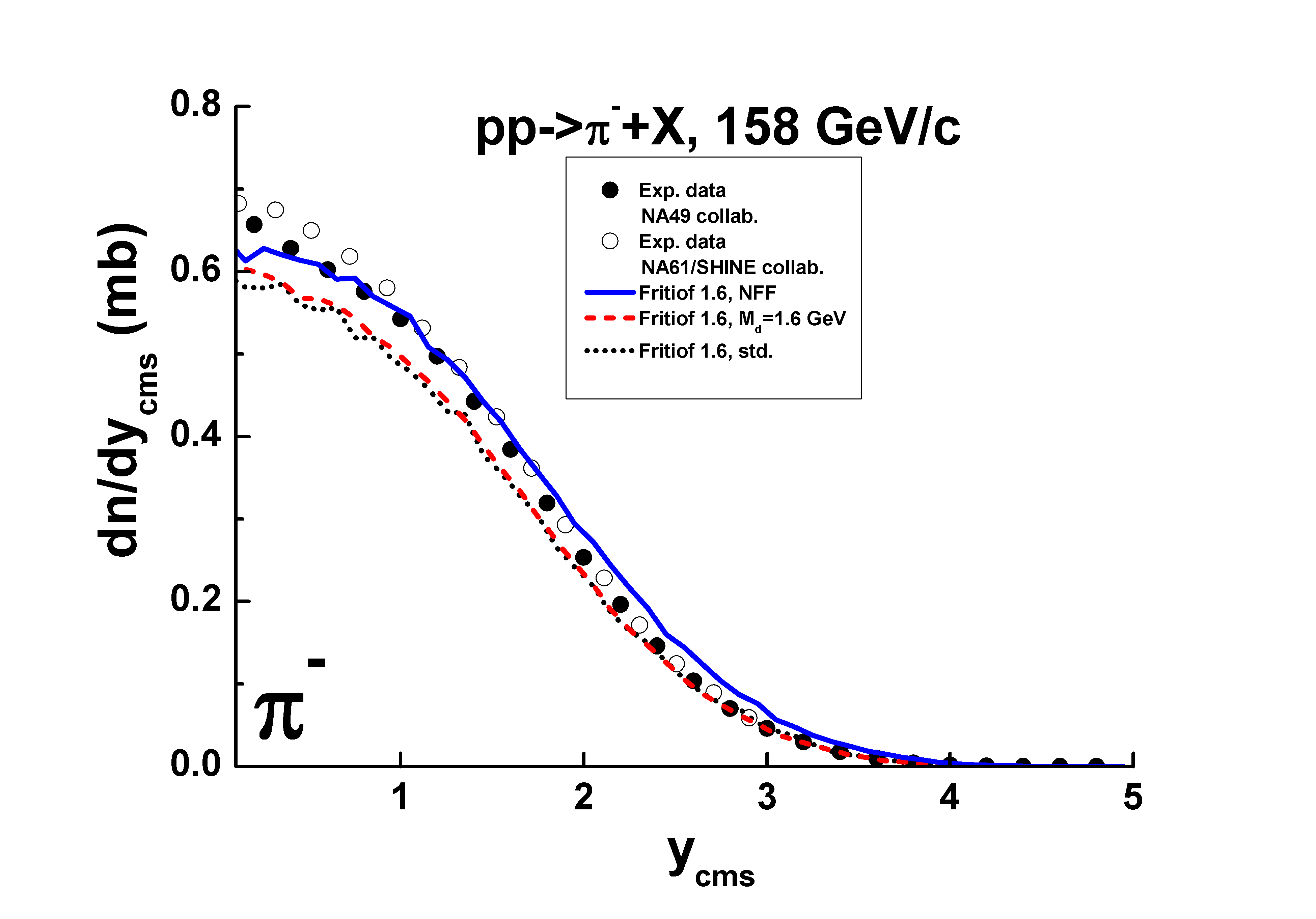}
\caption{Rapidity distributions of protons and $\pi^\pm$-mesons in $pp$ interactions at 158 GeV/c.
Closed points are the NA49 experimental data \protect{\cite{NA49pp2pi,NA49pp2p}}. Lines are
the Fritiof 1.6 model calculations (see text).}
\label{FriTuning}
\end{figure}
%\vspace{-3mm}

Tuning of the model at $P_{lab}=$ 158 GeV/c included 2 steps. At the first one, a probability
of the diffraction was enlarged increasing $M_d$ from 1.2 to 1.6 GeV (see dashed red lines in
Fig.~\ref{FriTuning}). At the second step, fragmentation functions of strings were changed,
especially the parameter "a" in the LUND string fragmentation function (see above) which
allowed to increase particle production in the central region.

A change of a string fragmentation function into baryons was more complicated. For this
the following lines were introduced in the subroutine LUZDIS of JETSET 6.3 code:
\begin{verbatim}
C...CHOICE OF Z, PREWEIGHTED FOR PEAKS AT LOW OR HIGH Z
      if(IFL1.gt.10) then                                    ! Uzhi
        Z=ZMAX*(RLU(0))**2                                   ! Uzhi
      else                                                   ! Uzhi
  100   Z=RLU(0)
...........................................
C...WEIGHTING ACCORDING TO CORRECT FORMULA
        IF(Z.LE.FB/(50.+FB).OR.Z.GE.1.) GOTO 100
        FVAL=(ZMAX/Z)*EXP(FB*(1./ZMAX-1./Z))
        IF(FA.GT.0.01) FVAL=((1.-Z)/(1.-ZMAX))**FA*FVAL
        IF(FVAL.LT.RLU(0)*FPRE) GOTO 100
      endif                                                  ! Uzhi
\end{verbatim}
The change operates at MST(10)=0. Results of the changes are presented in Fig.~\ref{FriTuning}
by solid blue lines. Dotted lines are standard model predictions.

It was checked that these allowed to describe $pp$-data at 20, 31, 40 and 80 GeV/c.

\subsection{Tuning of the UrQMD model}

\begin{figure}[cbth]
\includegraphics[width=75mm,height=30mm,clip]{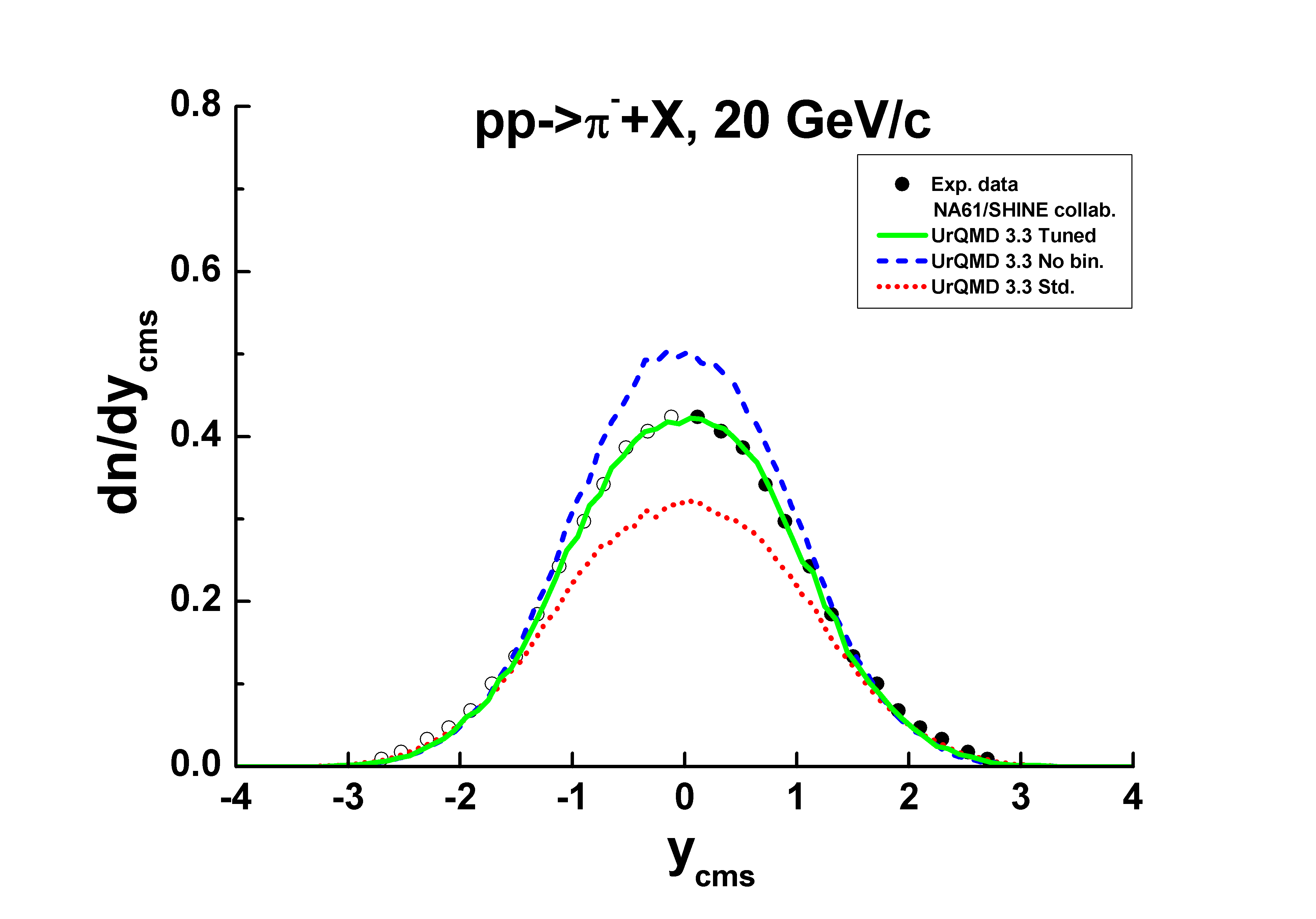}\includegraphics[width=75mm,height=30mm,clip]{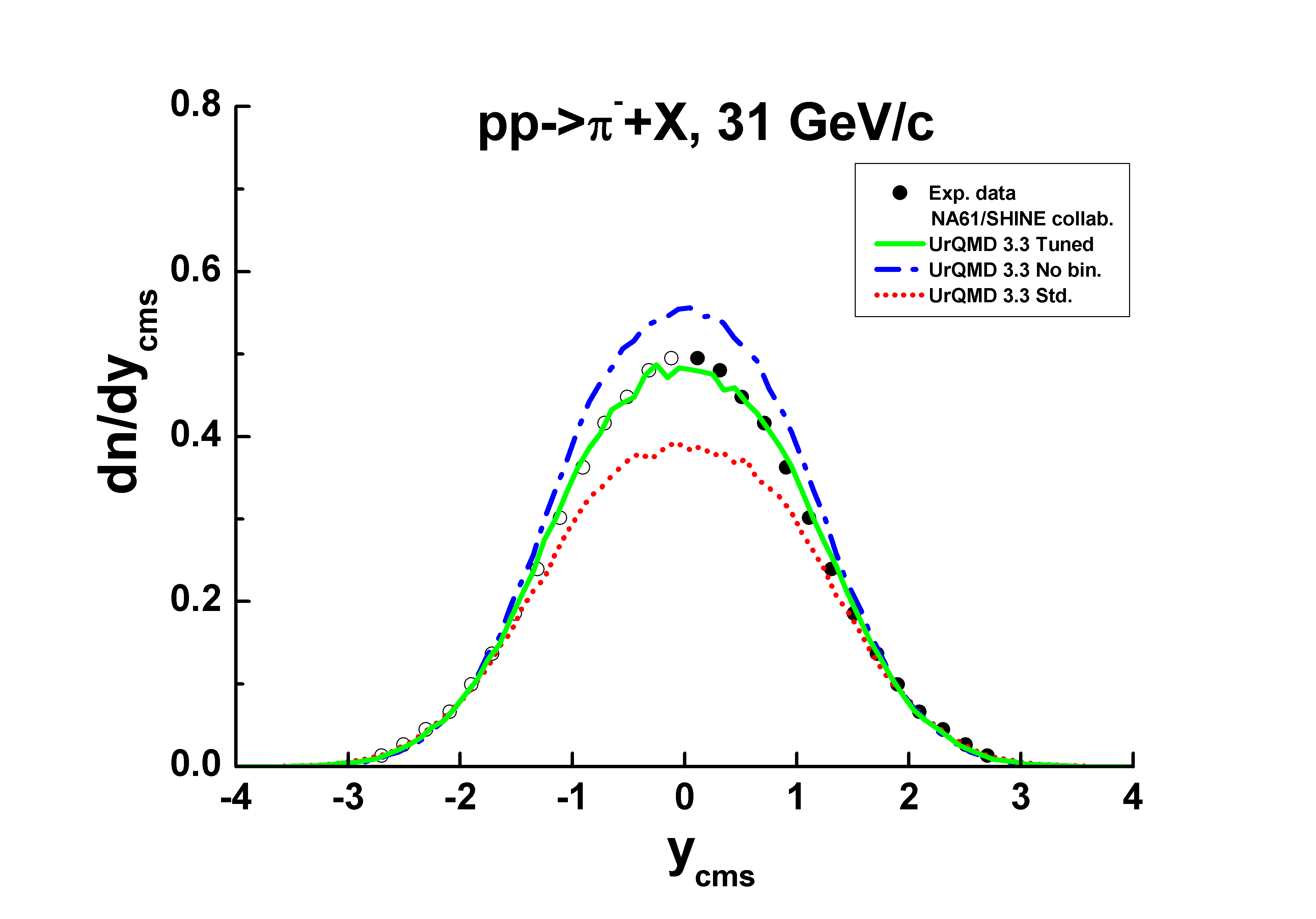}
\includegraphics[width=75mm,height=30mm,clip]{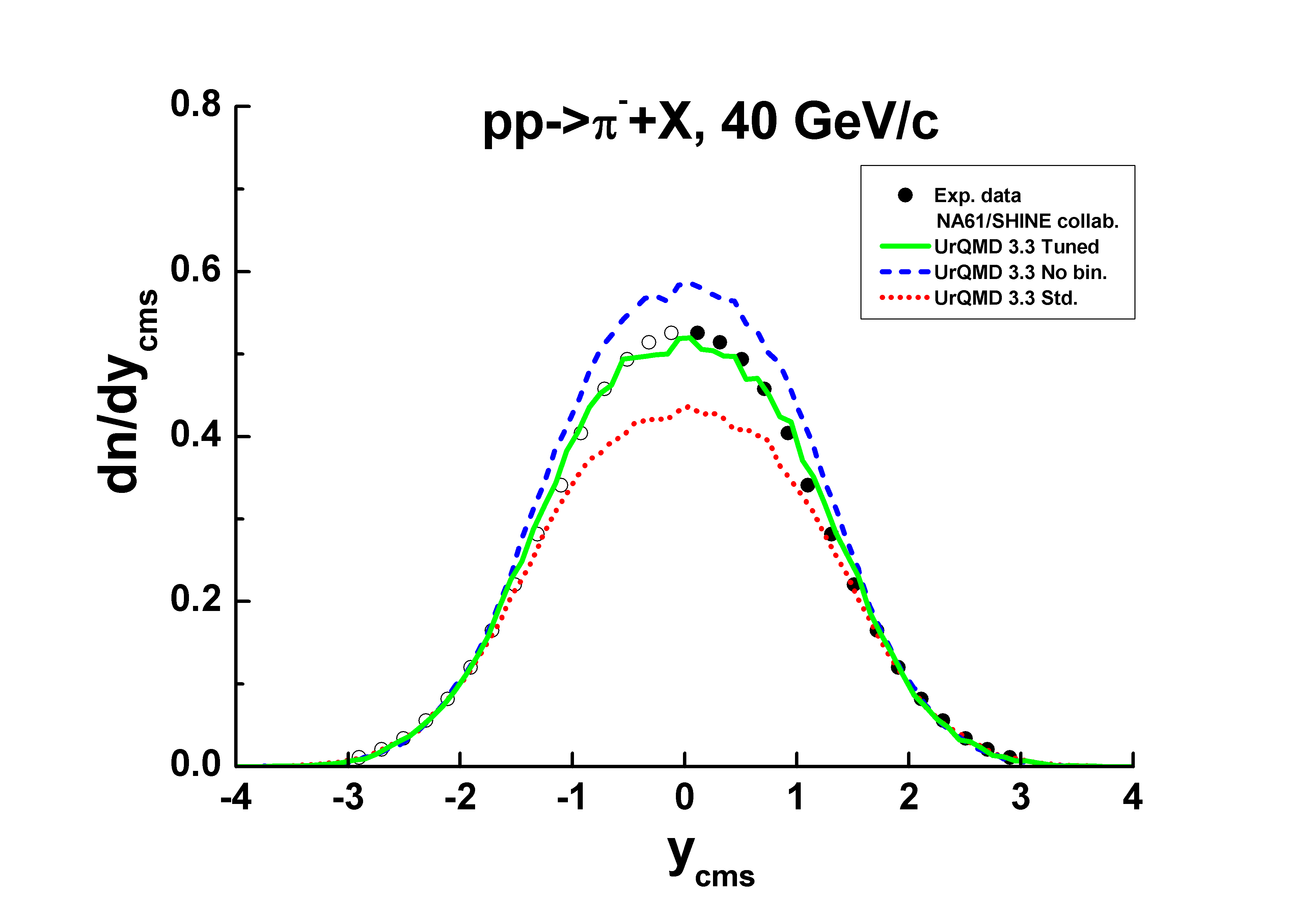}\includegraphics[width=75mm,height=30mm,clip]{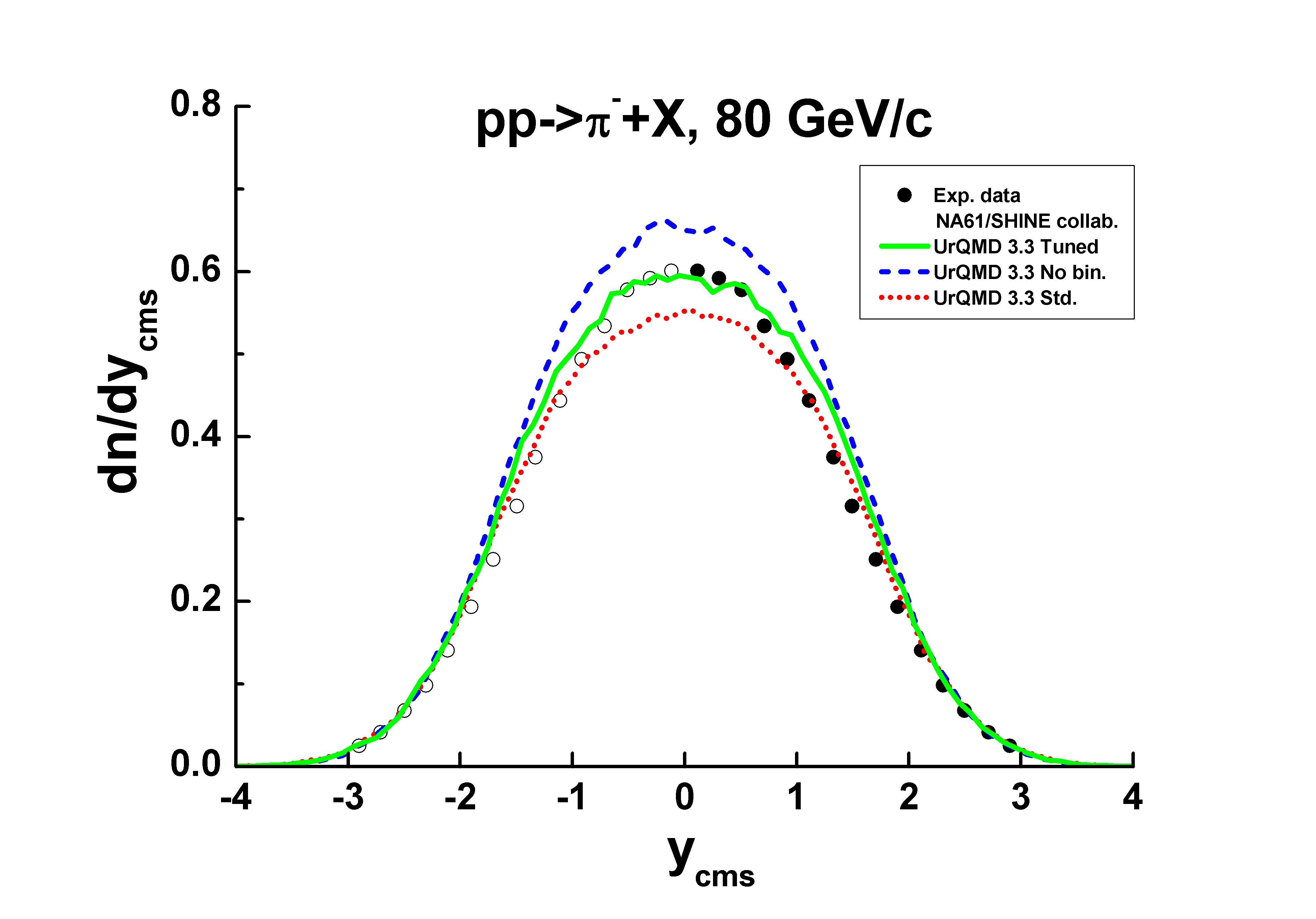}
\caption{Rapidity distributions of $\pi^-$-mesons in $pp$ interactions. Closed points
are the NA61/SHINE experimental data \protect{\cite{NA61}}, the open points are the data
reflected at mid-rapidity. Lines are UrQMD model calculations: solid lines are obtained
with tuned FTF process cross sections, long-dashed ones -- without the binary reactions,
short-dashed ones -- standard UrQMD model calculations.}
\label{NA61nobin}
\end{figure}

The accounting of the binary reactions is the main difference between the UrQMD model
and other Fritiof model implementations. Thus, one can suppose that an erasing\footnote{See
details in \protect{\cite{UzhiUrQ}}.}$^)$ of the reactions in the model eliminates the difference
between the model's predictions. Really, it is so, as it shown in Fig.~\ref{NA61nobin} where
short-dashed red lines show calculations by the original model, and long-dashed blue ones are calculations
without the binary reactions. As seen, the last calculations overestimate the data. The model
results can be easily improved introducing a probability of the Fritiof processes like
$1-1.1/\sqrt{s}$ (see solid green lines in Fig.~\ref{NA61nobin}). These indicate that the binary
reaction cross sections are not tuned quite well in the model.

%\vspace{-3mm}

The probability of the Fritiof processes was introduced in the file scatter.f of the model as:
\begin{verbatim}
         call normit (sigma,isigline)

      if((ityp1.le.100).and.(ityp2.le.100).and.(sqrts.ge.3.5d0)) then ! Uzhi
        UzhiXin=sigma(0)-sigma(1)                                     ! Uzhi
        UzhiSum=UzhiXin-sigma(9)                                      ! Uzhi
        sigma(9)=UzhiXin*(1.-1.1/SqrtS)                               ! Uzhi
        UzhiFac=(UzhiXin-sigma(9))/UzhiSum                            ! Uzhi
        do ii=2,nCh-1                                                 ! Uzhi
          sigma(ii)=sigma(ii)*UzhiFac                                 ! Uzhi
        enddo                                                         ! Uzhi
      endif                                                           ! Uzhi
\end{verbatim}

\begin{figure}[cbth]
\includegraphics[width=50mm,height=45mm,clip]{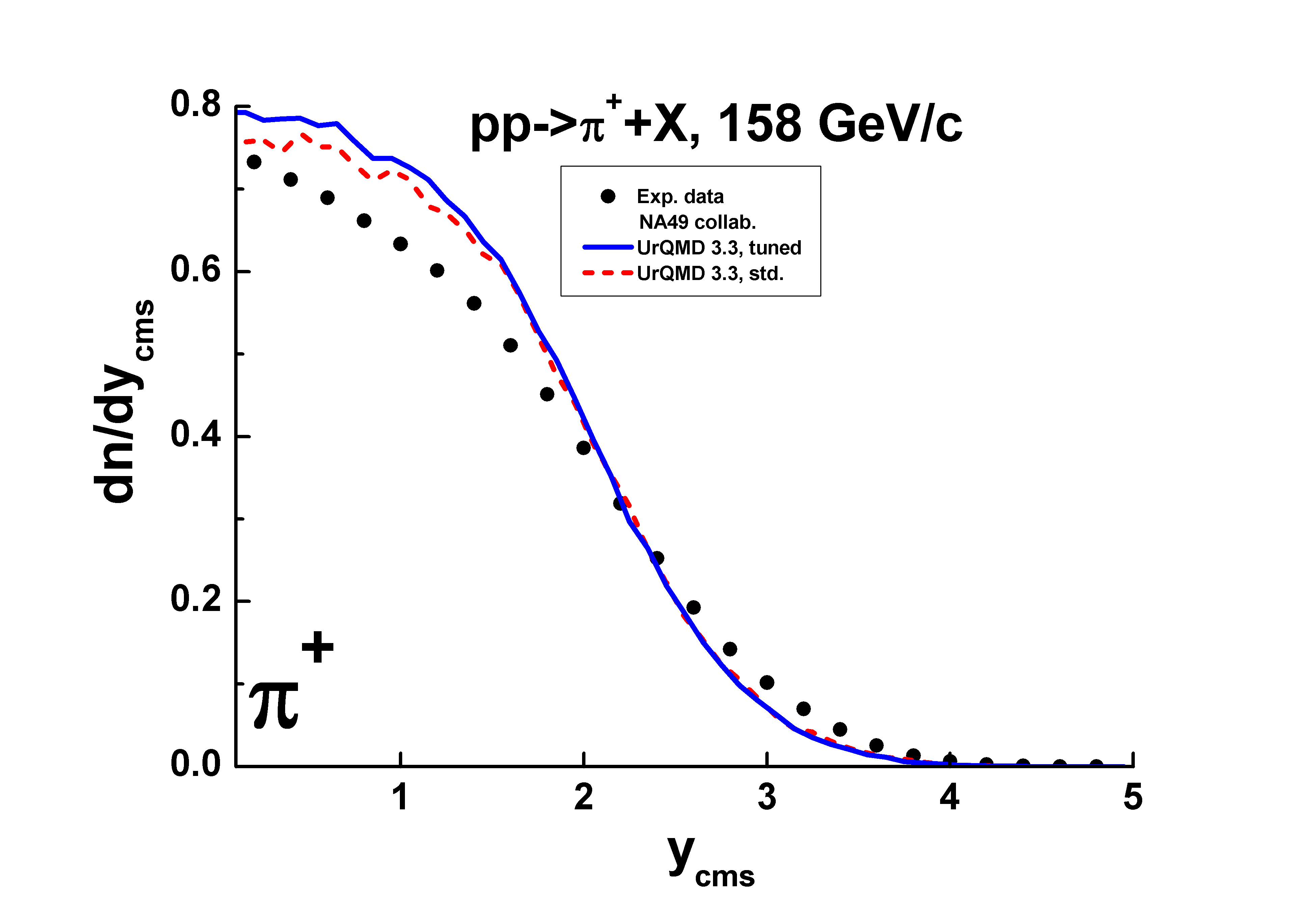}\includegraphics[width=50mm,height=45mm,clip]{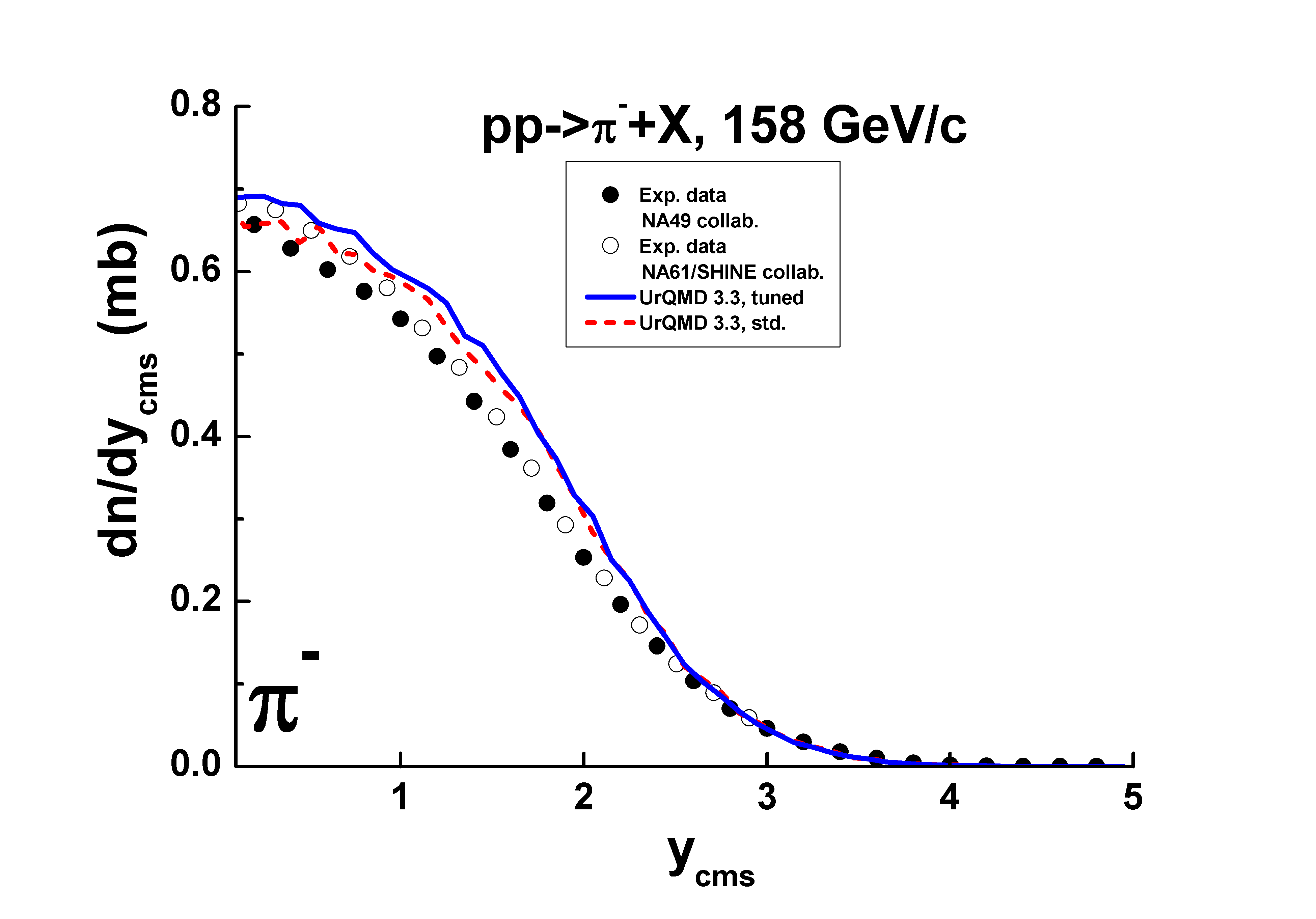}\includegraphics[width=50mm,height=45mm,clip]{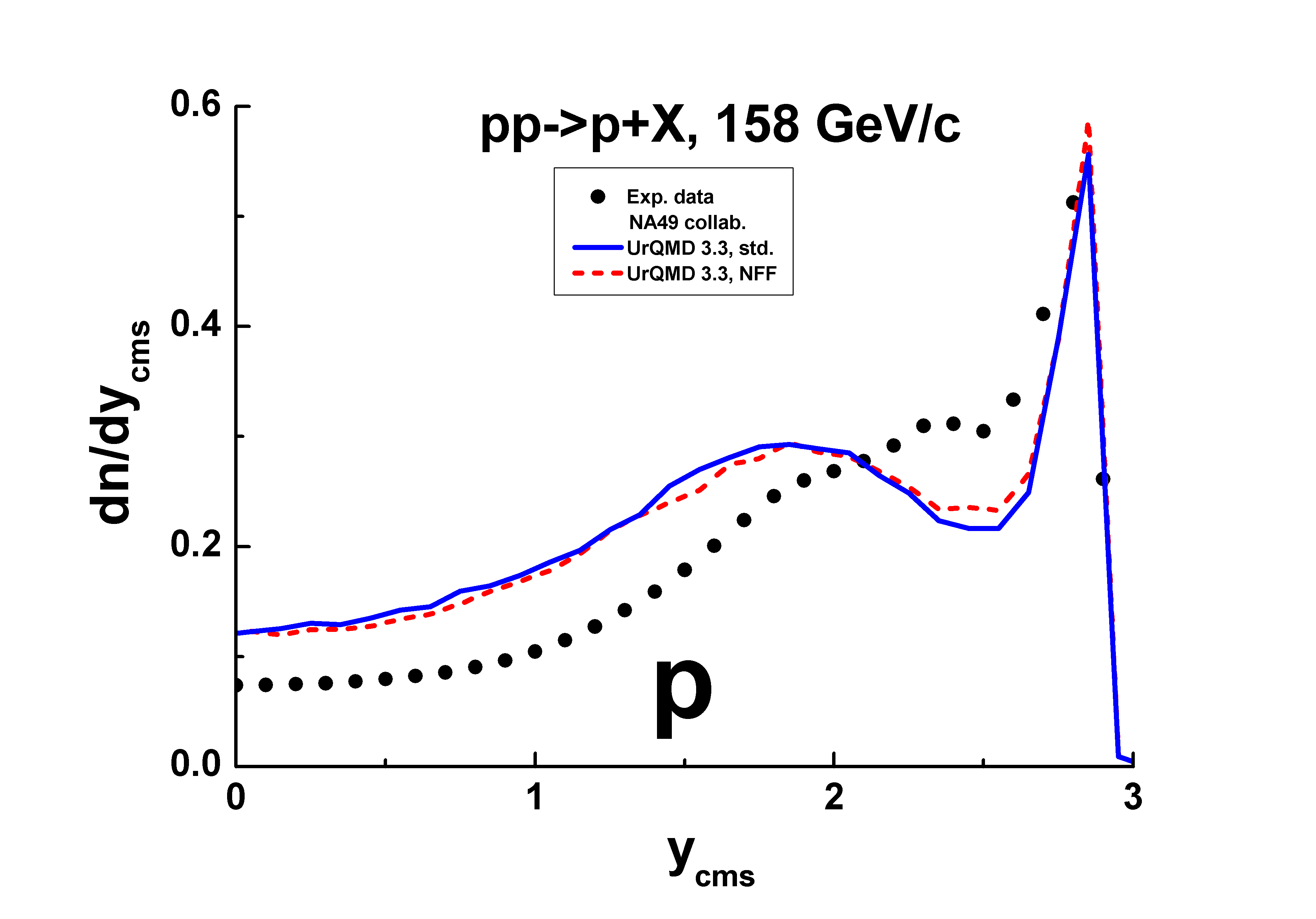}
\caption{Rapidity distributions of protons and $\pi^\pm$-mesons in $pp$ interactions at 158 GeV/c.
Closed points are the NA49 experimental data \protect{\cite{NA49pp2pi,NA49pp2p}}. Lines are
the UrQMD model calculations (see text).}
\label{UrTuning}
\end{figure}

Results of the tuning for $pp$-interactions at 158 GeV/c are presented in Fig.~\ref{UrTuning}
by solid lines. Dashed lines are the standard model calculations. As seen, the model reproduces
correctly $\pi^-$ meson production at $y_{cms}\sim$ 0, but there is an essential difference
between the data and the model calculations at other $y_{cms}$. The difference becomes larger
for $\pi^+$ spectra.
%\vspace{-3mm}

As seen also, the tuning does not affect on the proton spectra. Attempts to improve
the description of the proton spectrum were not successful.

\section{Verification of diffraction dissociation simulations}
The above mentioned differences of the model implementations lead to various
predictions of the diffraction peak height shown in Fig.~\ref{pp158prot}. To understand
the results, let us look at model calculations and experimental data at various energies.

\begin{figure}[cbth]
    \begin{center}
        \includegraphics[width=80mm,height=33mm,clip]{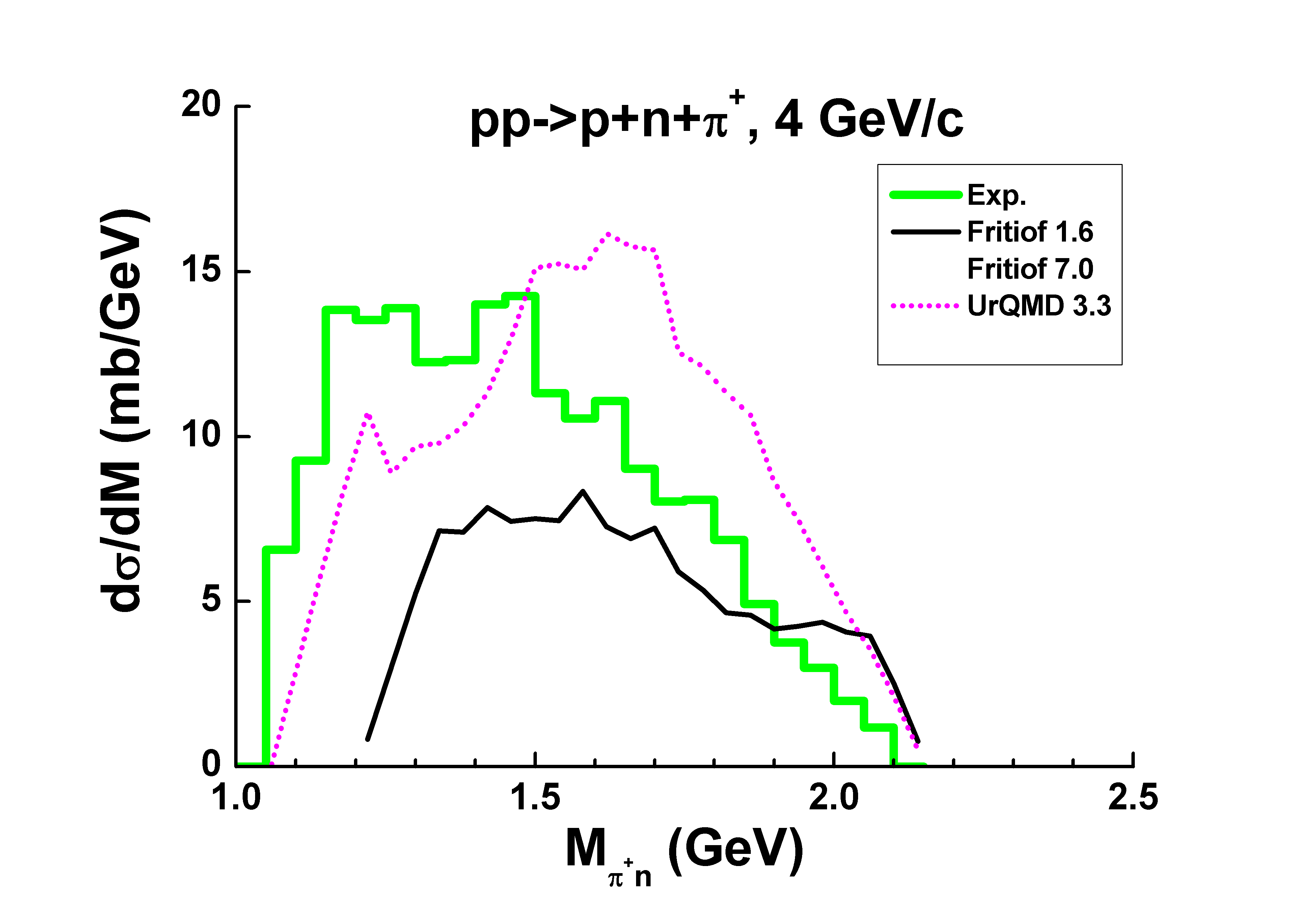}\includegraphics[width=80mm,height=33mm,clip]{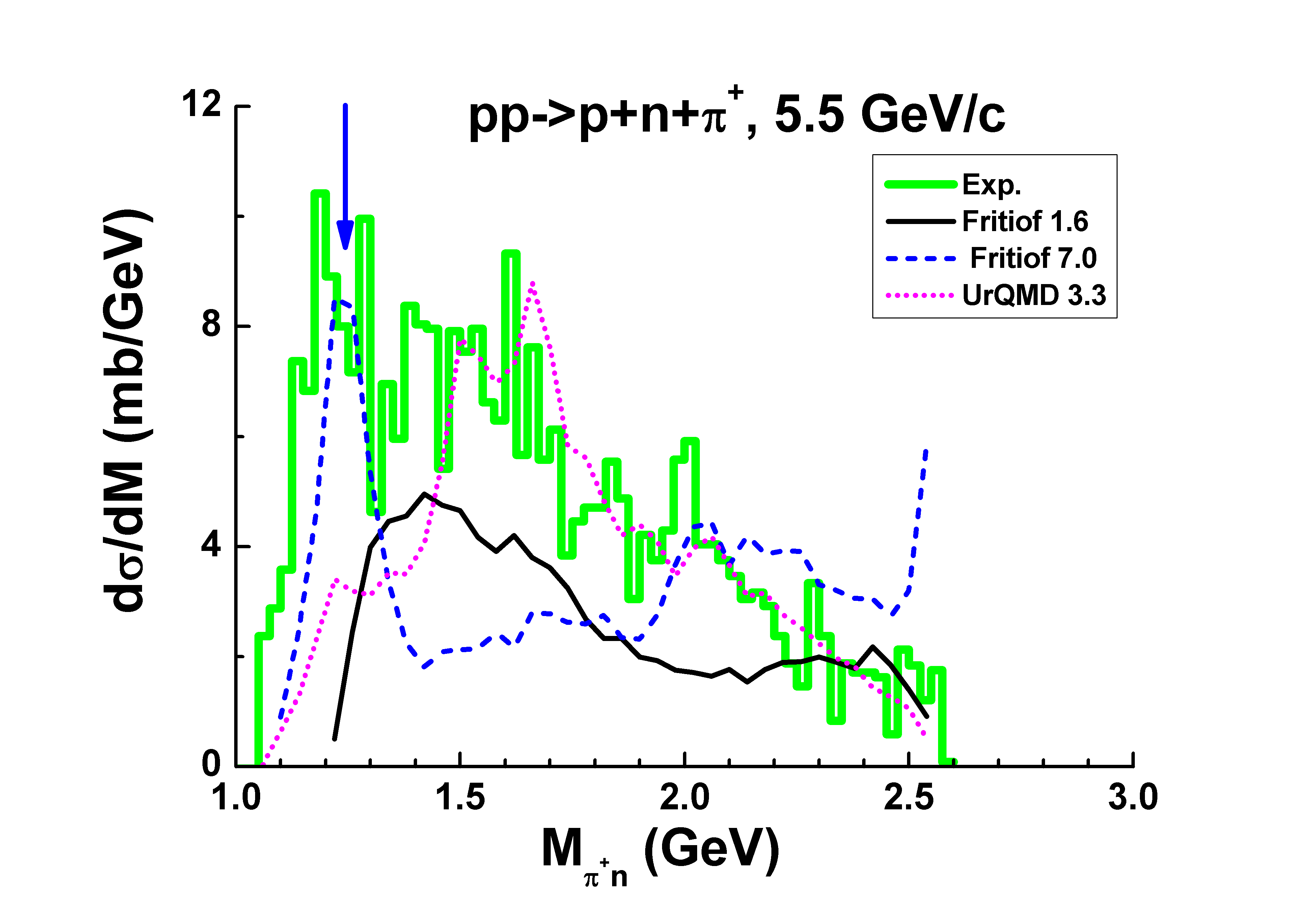}
        \includegraphics[width=80mm,height=33mm,clip]{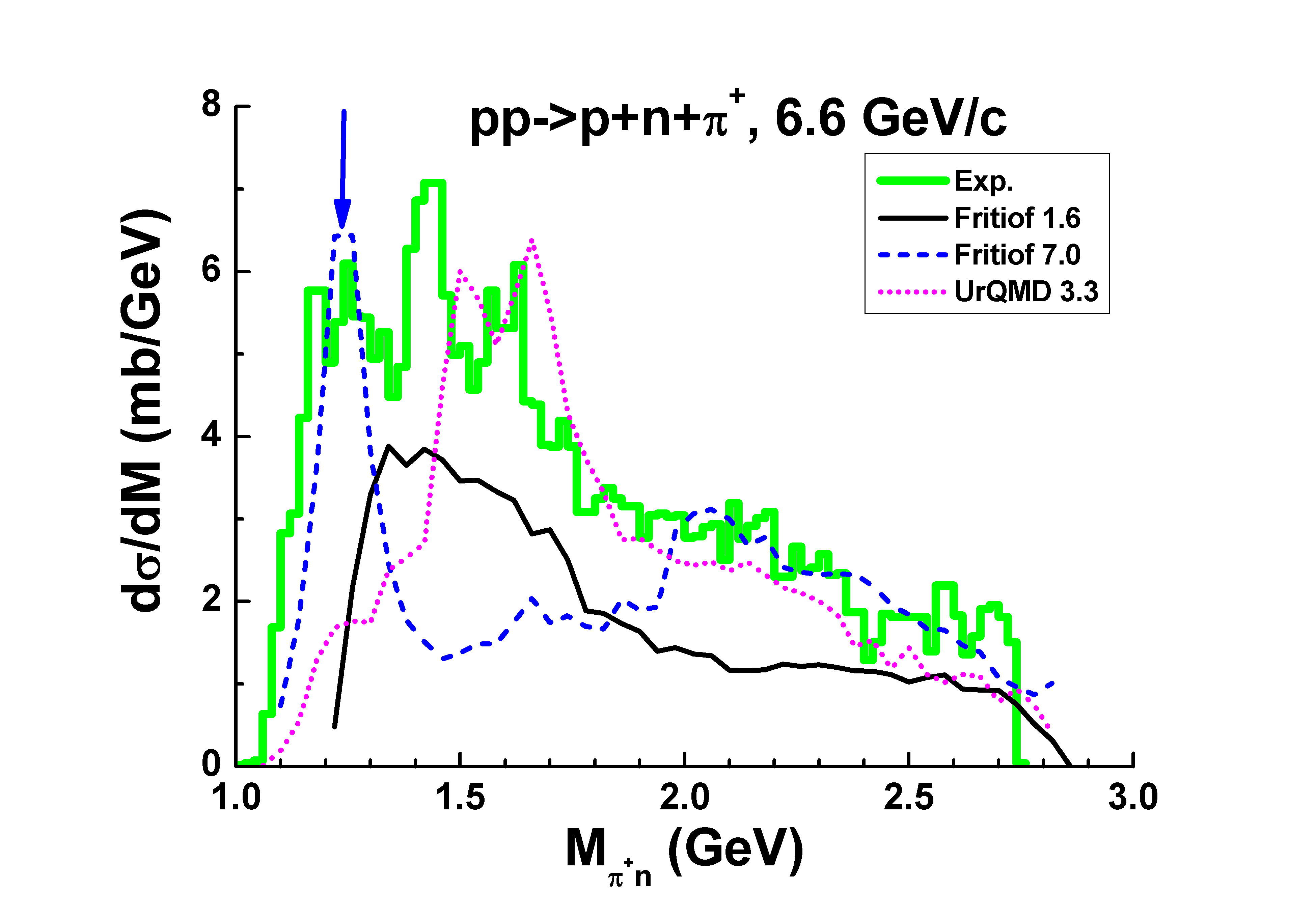}\includegraphics[width=80mm,height=33mm,clip]{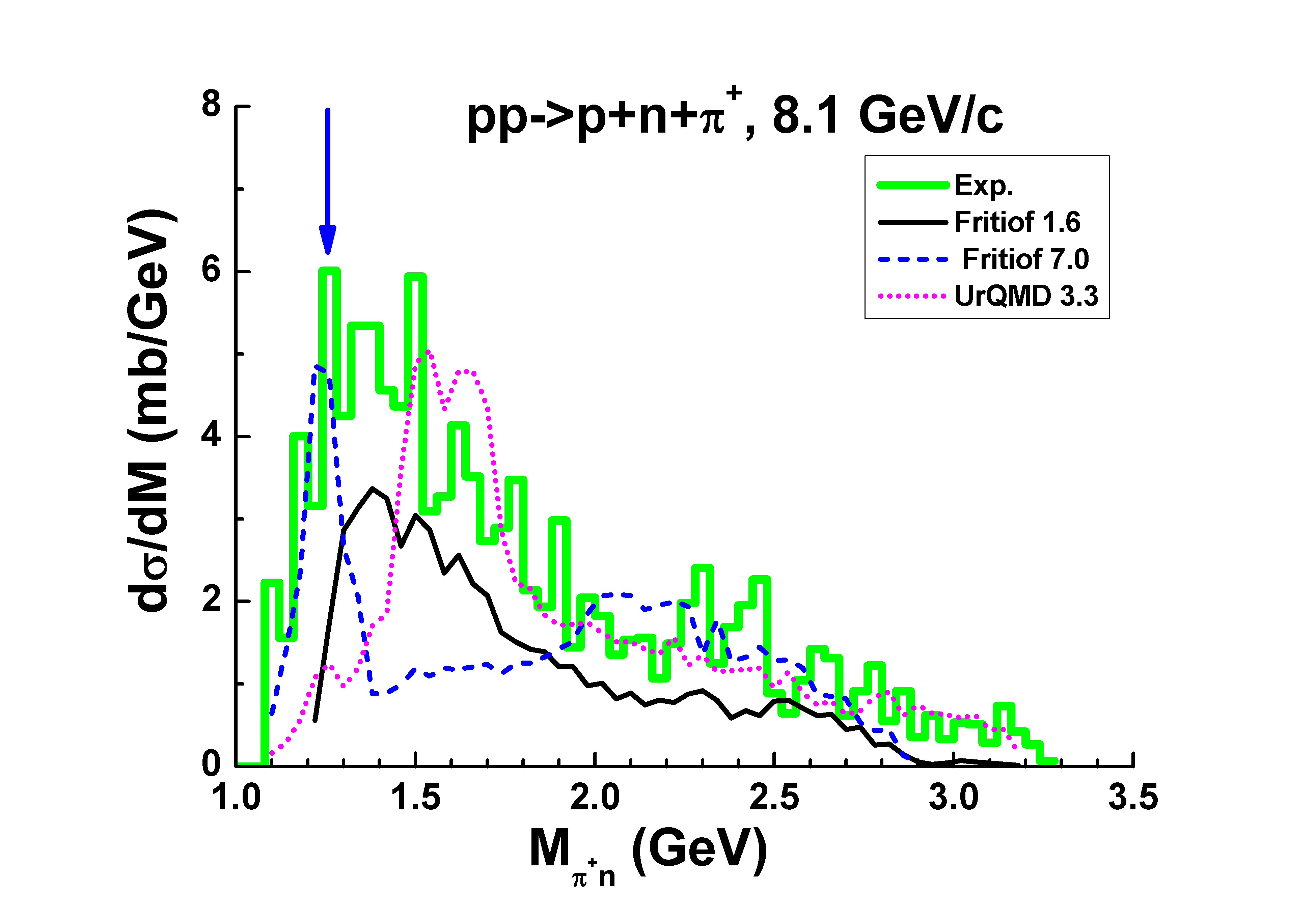}
        \includegraphics[width=80mm,height=33mm,clip]{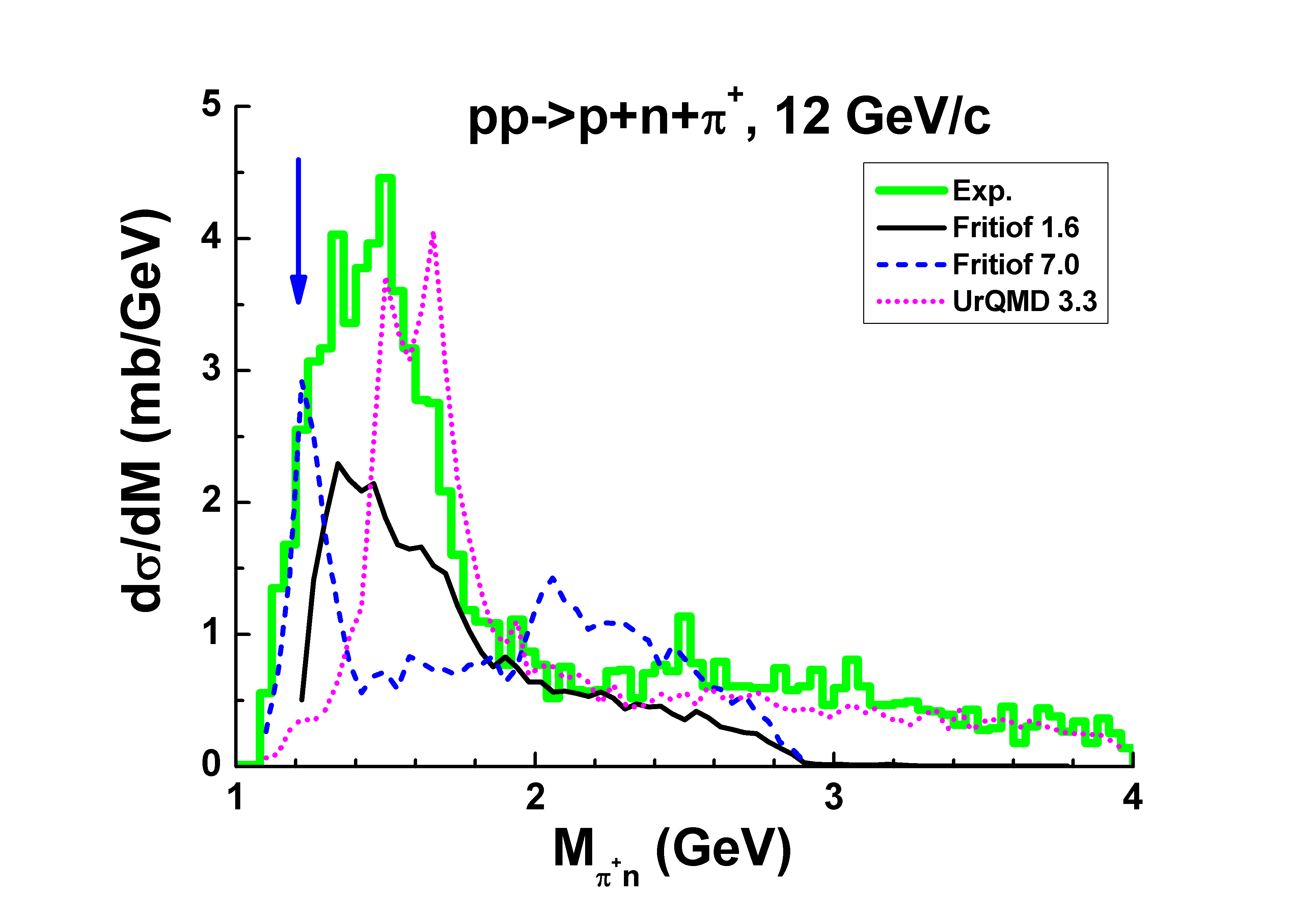}\includegraphics[width=80mm,height=33mm,clip]{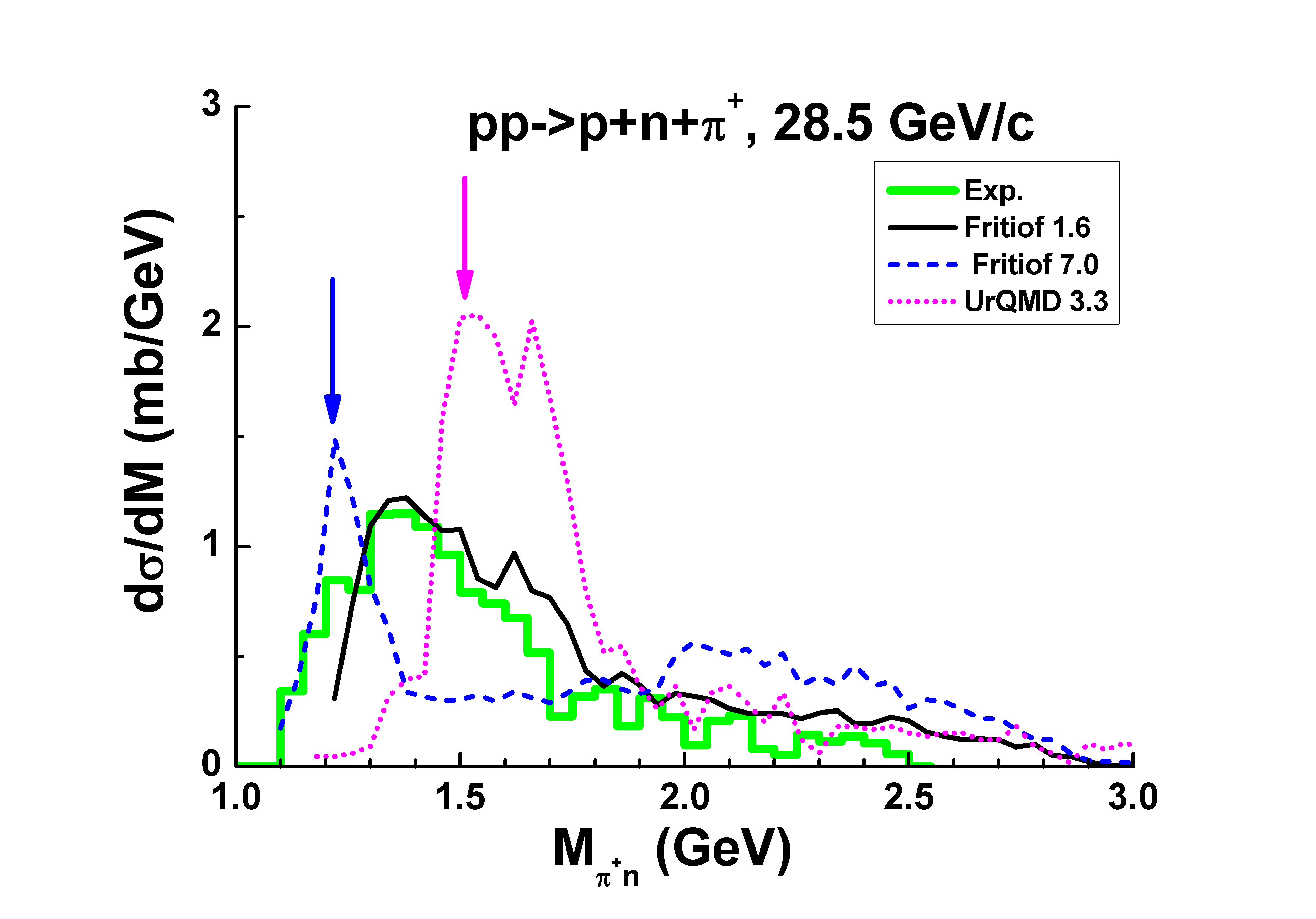}
        \caption{Mass distributions of $n\pi^+$ pairs in the reactions
                 $pp\rightarrow p+n+\pi^+$. Points are experimental data
                 presented in \protect{\cite{P4,P5_5,P6_6,P8_1,P12,P28_5}}.
                 Lines are model calculations.}
        \label{pnPip}
    \end{center}
\end{figure}

\begin{figure}[cbth]
    \begin{center}
        \includegraphics[width=80mm,height=33mm,clip]{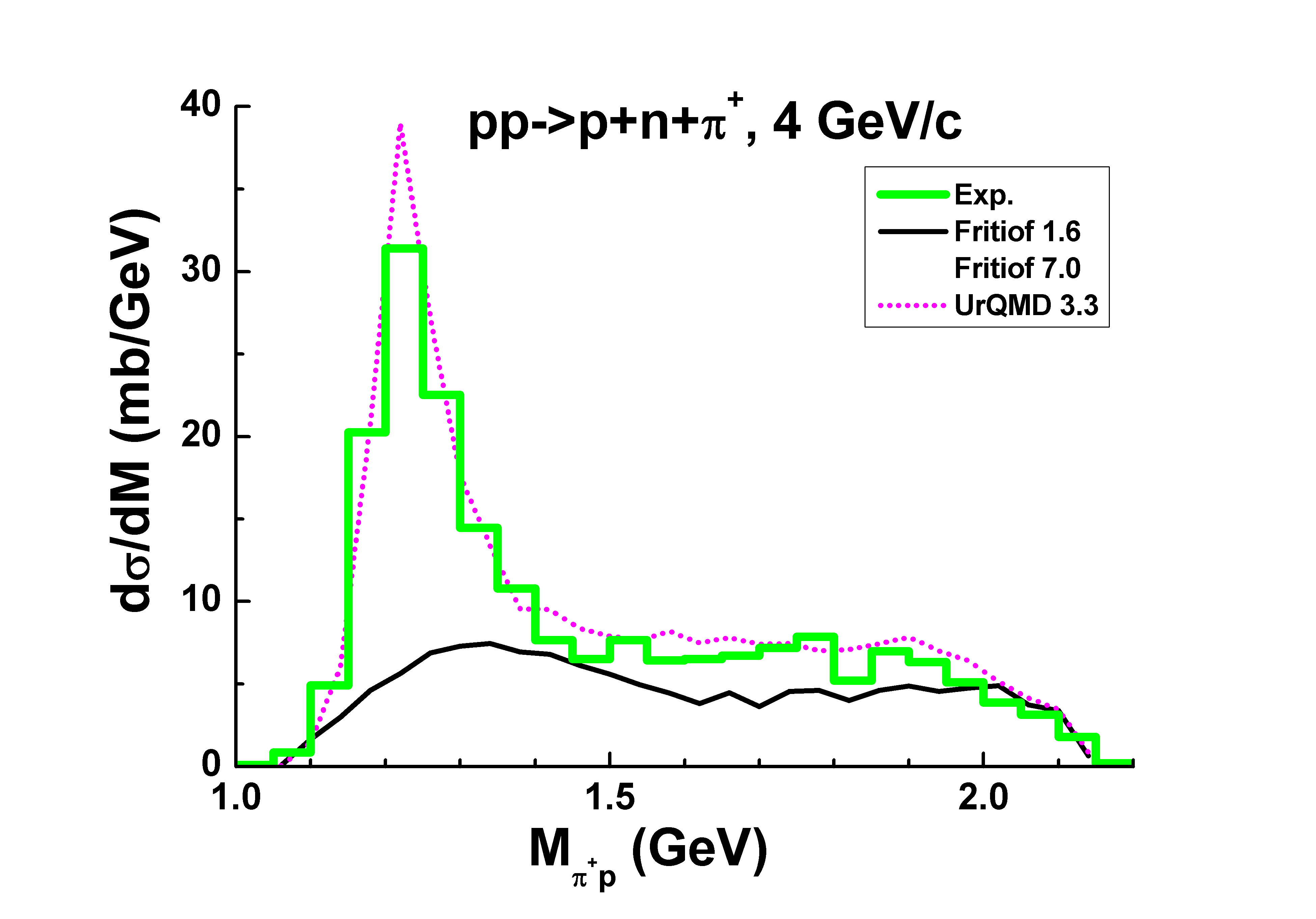}\includegraphics[width=80mm,height=33mm,clip]{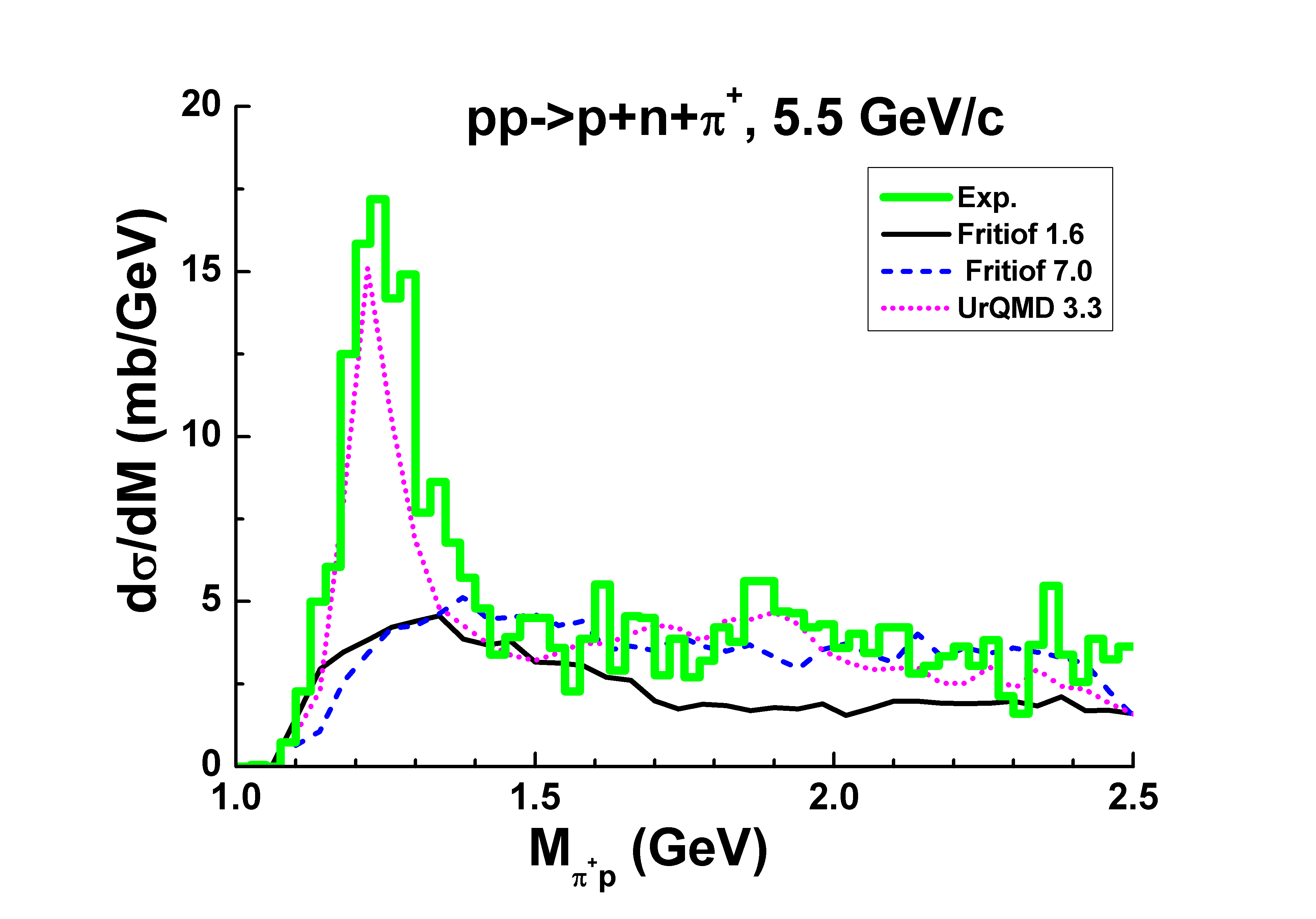}
        \includegraphics[width=80mm,height=33mm,clip]{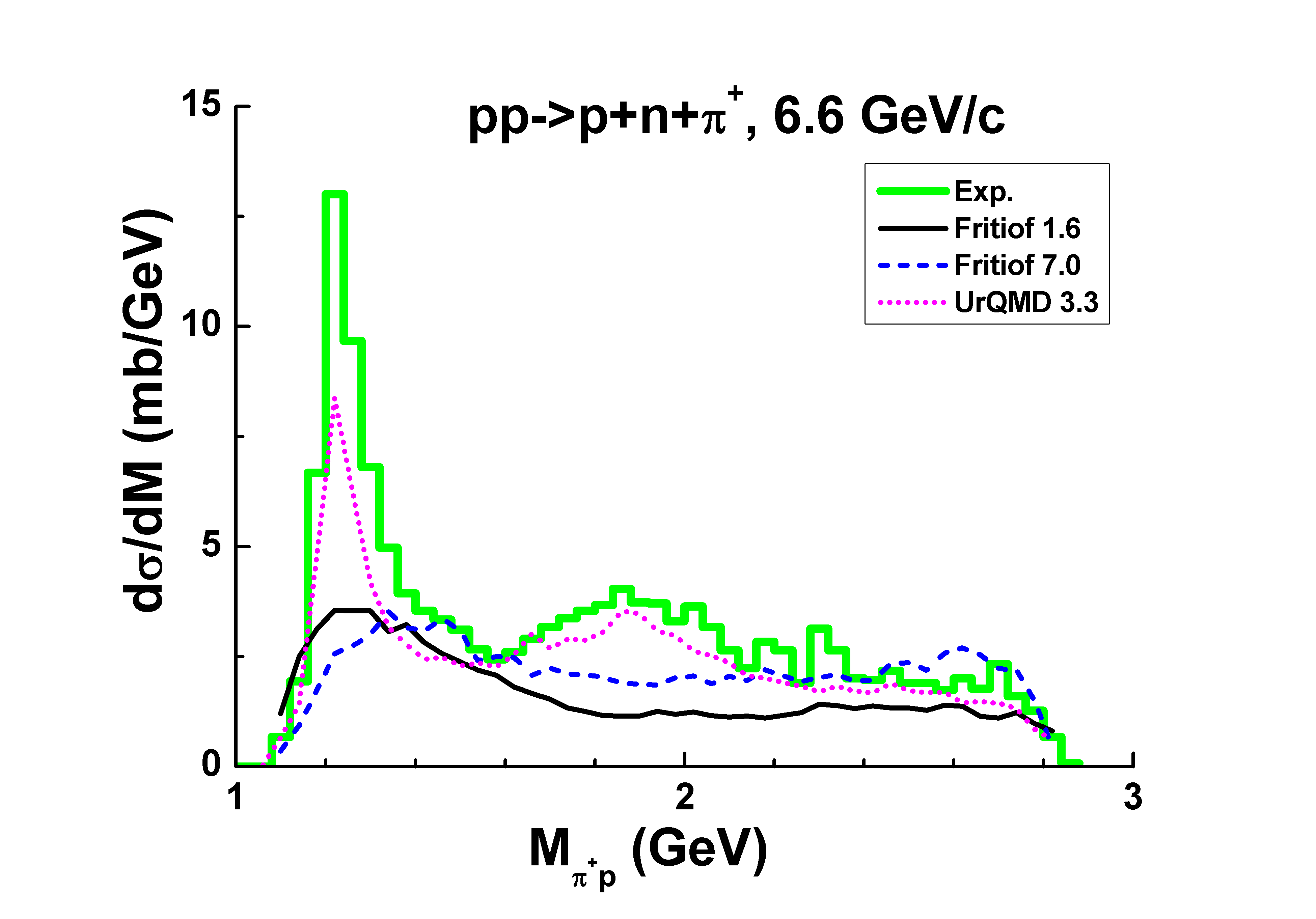}\includegraphics[width=80mm,height=33mm,clip]{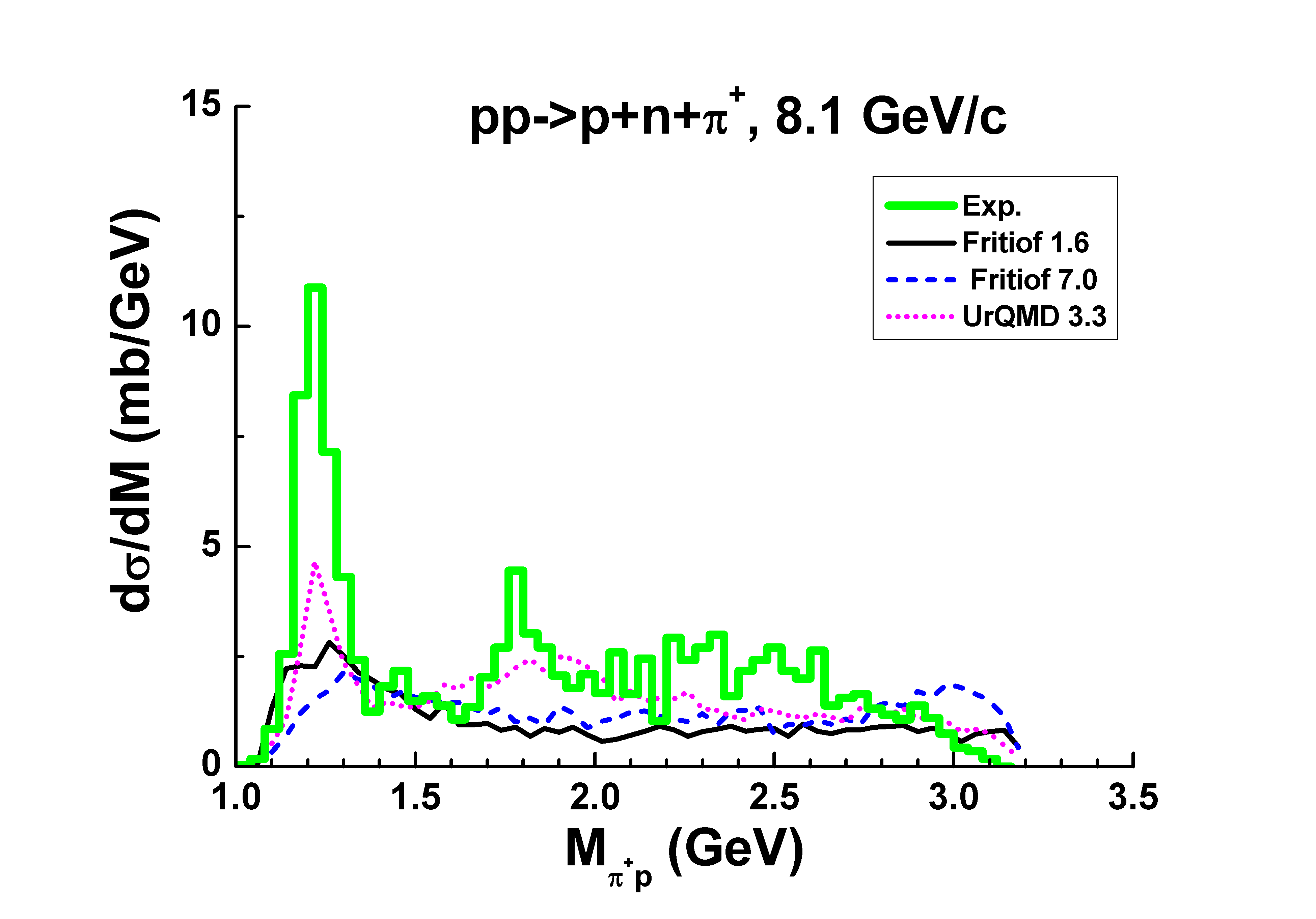}
        \includegraphics[width=80mm,height=33mm,clip]{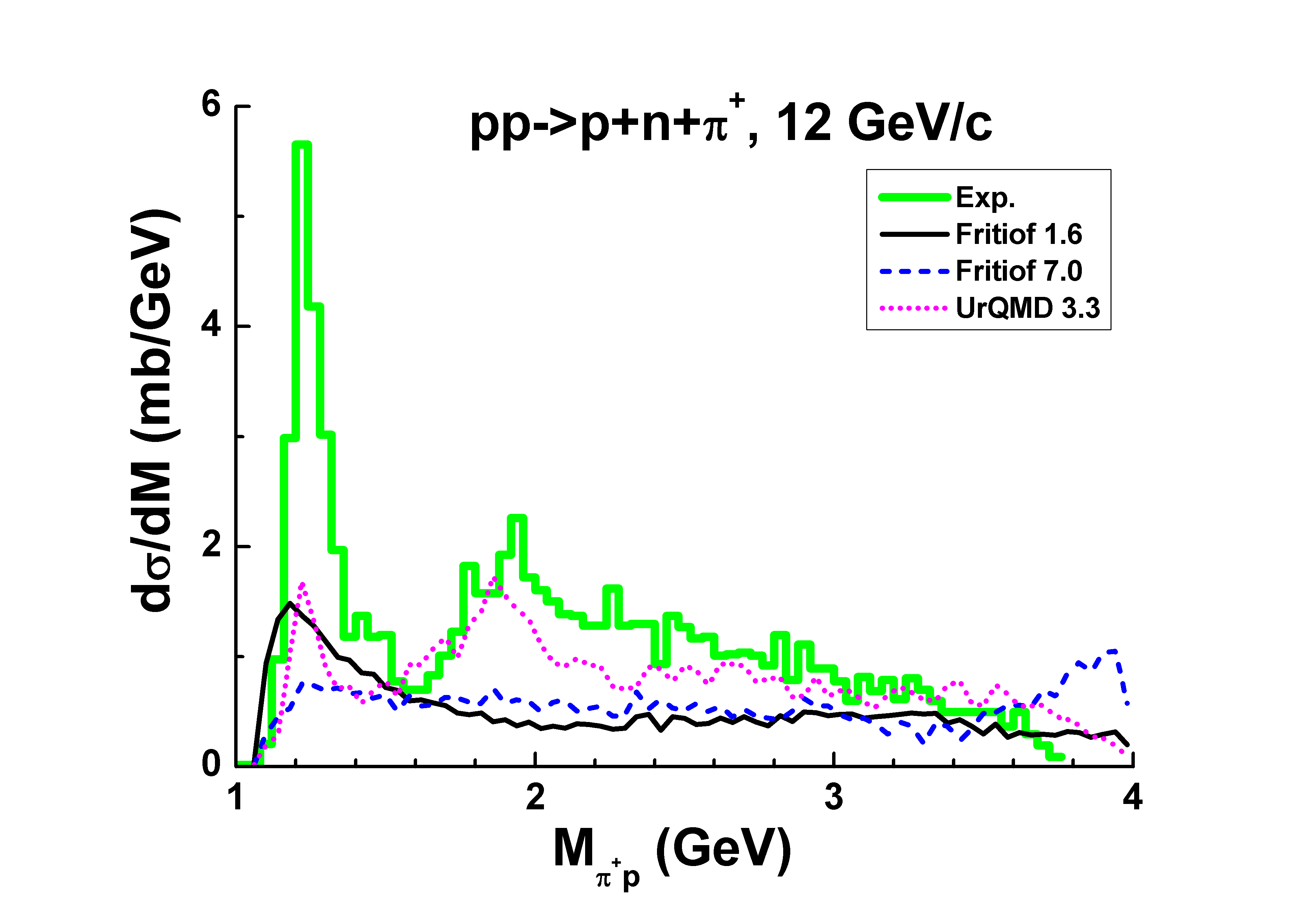}\includegraphics[width=80mm,height=33mm,clip]{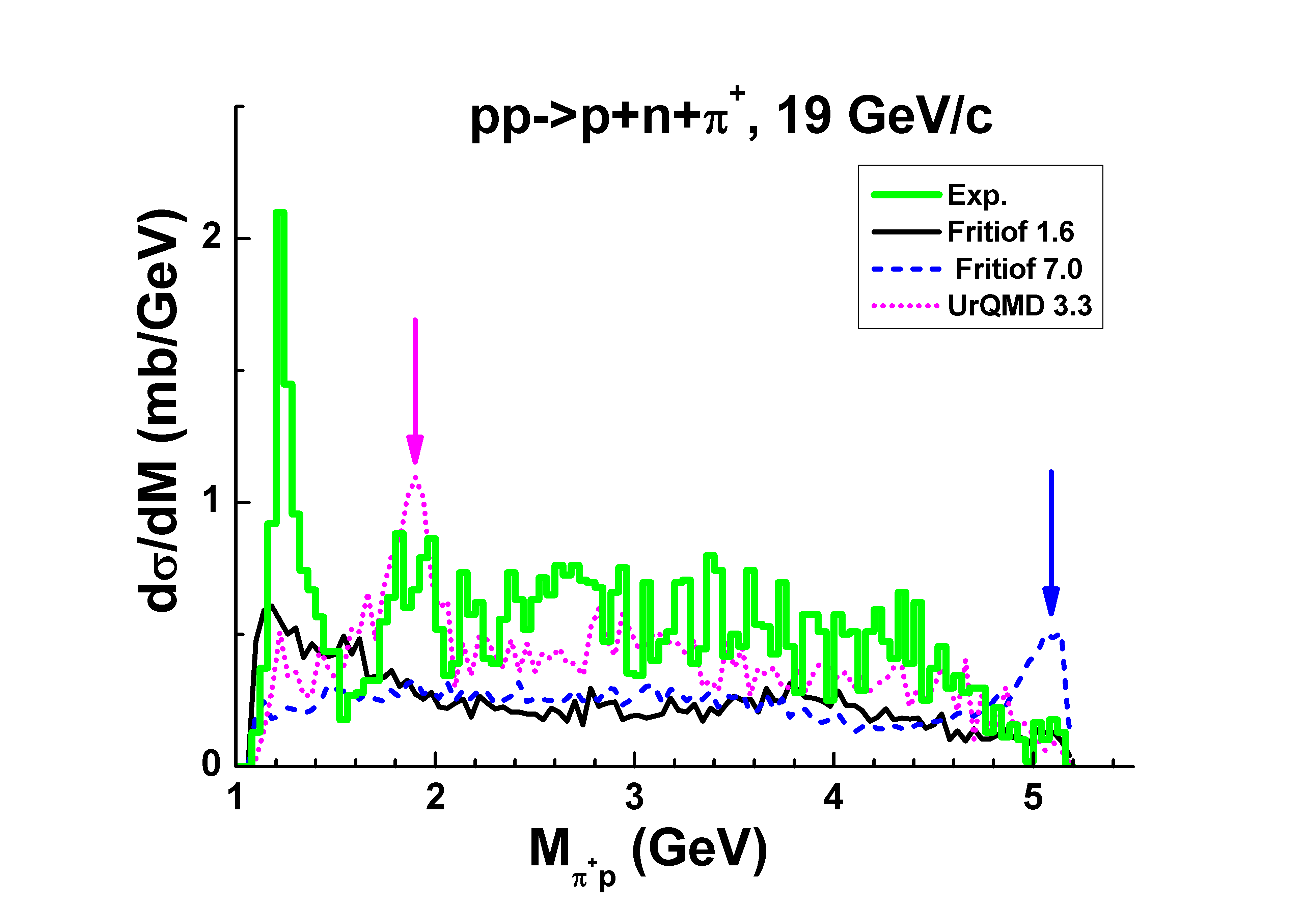}
        \caption{Mass distributions of $p\pi^+$ pairs in the reactions
                 $pp\rightarrow p+n+\pi^+$. Points are experimental data presented in
                 \protect{\cite{P4,P5_5,P6_6,P8_1,P12,P19}}.
                 Lines are model calculations.}
        \label{pPipn}
    \end{center}
\end{figure}

There are a lot of data at low and high energies. Some of them are shown in
Fig.~\ref{pnPip} and \ref{pPipn} together with model calculations. The diffraction
dissociation in a simple case can be seen in the reactions:
$p+p\rightarrow p+p'\rightarrow p+(p\pi^0)$, $p+p\rightarrow p+p'\rightarrow p+(n\pi^+)$.
Of course, there can be other reactions with non-vacuum exchanges in the $t$-channel:
\begin{equation} p+p\rightarrow p+\Delta^+,~N^*\rightarrow p+(p\pi^0),\label{pppi0}\end{equation}
\begin{equation} p+p\rightarrow p+\Delta^+,~N^*\rightarrow p+(n\pi^+),\label{pnpip}\end{equation}
\begin{equation} p+p\rightarrow n+\Delta^{++}\rightarrow n+(p\pi^+).\label{nDpp}\end{equation}
The reactions \ref{pppi0}, \ref{pnpip}, \ref{nDpp} are absent in the original Fritiof
model. Thus, mass spectra of various systems in the reaction
$p+p\rightarrow  p+n+\pi^+$ are smooth enough according to the Fritiof 1.6 implementation.

Another situation takes place in the Fritiof 7.0. Final states of diffraction
dissociations in the implementation typically are $p+q+qq$. The LUND string fragmentation
algorithm interprets them as $p+\Delta^+(1232)$ systems if their masses are sufficiently
small. Thus the corresponding peak is observed in calculated $M_{n\pi^+}$ mass spectra
(see Fig.~\ref{pnPip}). Due to this also, the form of the spectra are not agree with
experimental ones at $M_{n\pi^+}\leq 2$ GeV.

According to the UrQMD model, processes with a creation of $\Delta^+(1232)$ or
$N^{*+}(1440)$ isobars are not dominating in the reactions. Instead, $N^{*+}$ isobars
with masses 1520 -- 1700 MeV are copiously produced especially at high energies
in a disagreement with the experimental data, see Fig.~\ref{pnPip}. Usually,
a kinematical peak is seen in experimental high energy data at $M_x \sim$ 1400 MeV.
It is not reproduced in the models.
Resonances, especially
the Roper resonance, are not responsible for the structure of the peak in the
low mass region. There were a lot of papers devoted to its description.
Mainly they were done within the One-Pion-Exchange model taken its origin from
the well-known papers \cite{DrellHiida,Deck}. Many interesting results were obtained in
the model, and one can hope that they can be used in Monte Carlo event generators.

Mass distributions of $p\pi^+$ pairs in the reactions $pp\rightarrow p+n+\pi^+$
give information about non-vacuum exchanges. As seen in Fig.~\ref{pPipn}, peaks
connected with creations of $\Delta^{++}(1232)$-isobars are presented in
the experimental data. The UrQMD model reproduces the peaks at $p_{lab} <$ 6 GeV/c,
but it assumes that the yield of the resonances decreases very fast with energy
growth. Fritiof 1.6 and Fritiof 7.0 models predict smooth distributions without peaks.

Summing up, one can conclude that a description of the mass distributions in
the low mass region is a problem in all Fritiof-based model especially at high energies.

Experimental mass distributions at high energies are usually presented at various
values of 4-momen\-tum transfers, $t$. For their analysis one needs to reproduce
a $t$-dependence of the spectra which requires a lot of work because a slope of
a distribution on $t$ depends on a produced mass and on an energy of collisions.
The models do not assume such dependencies, as it is seen in Fig.~\ref{Tdep} where
experimental data \cite{Scham} are presented in a comparison with calculations results.

As seen also, the Fritiof 1.6 model does not describe the data at small parameter
$P_T=283$ MeV. At larger parameter, the slope of the calculated curves becomes
close to the experimental one. Now it is clearly seen, that the model underestimates
the diffraction cross sections.

The UrQMD model gave at the beginning very strange results: the calculated distributions
had step-like behaviour. To improve the model the following change has been done in
the model (file angdis.f):
\begin{verbatim}
c for jmax=12 the accuracy is better than 0.1 degree
c
*      do j=1,12 ! accuracy 2**-jmax    ! Uzhi
      do j=1,24  ! accuracy 2**-jmax    ! Uzhi
\end{verbatim}

\begin{figure}[cbth]
    \begin{center}
        \includegraphics[width=80mm,height=35mm,clip]{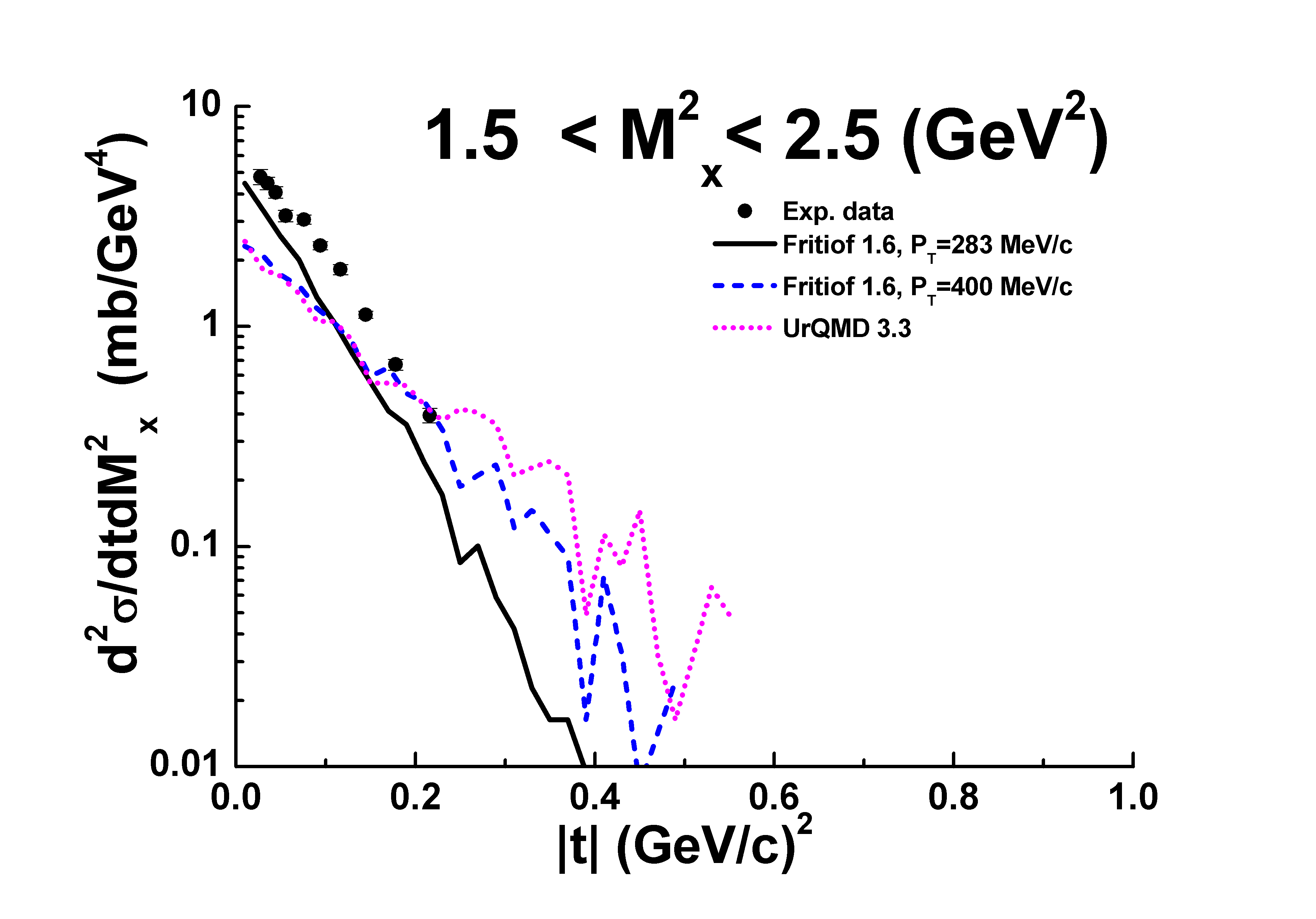}\includegraphics[width=80mm,height=35mm,clip]{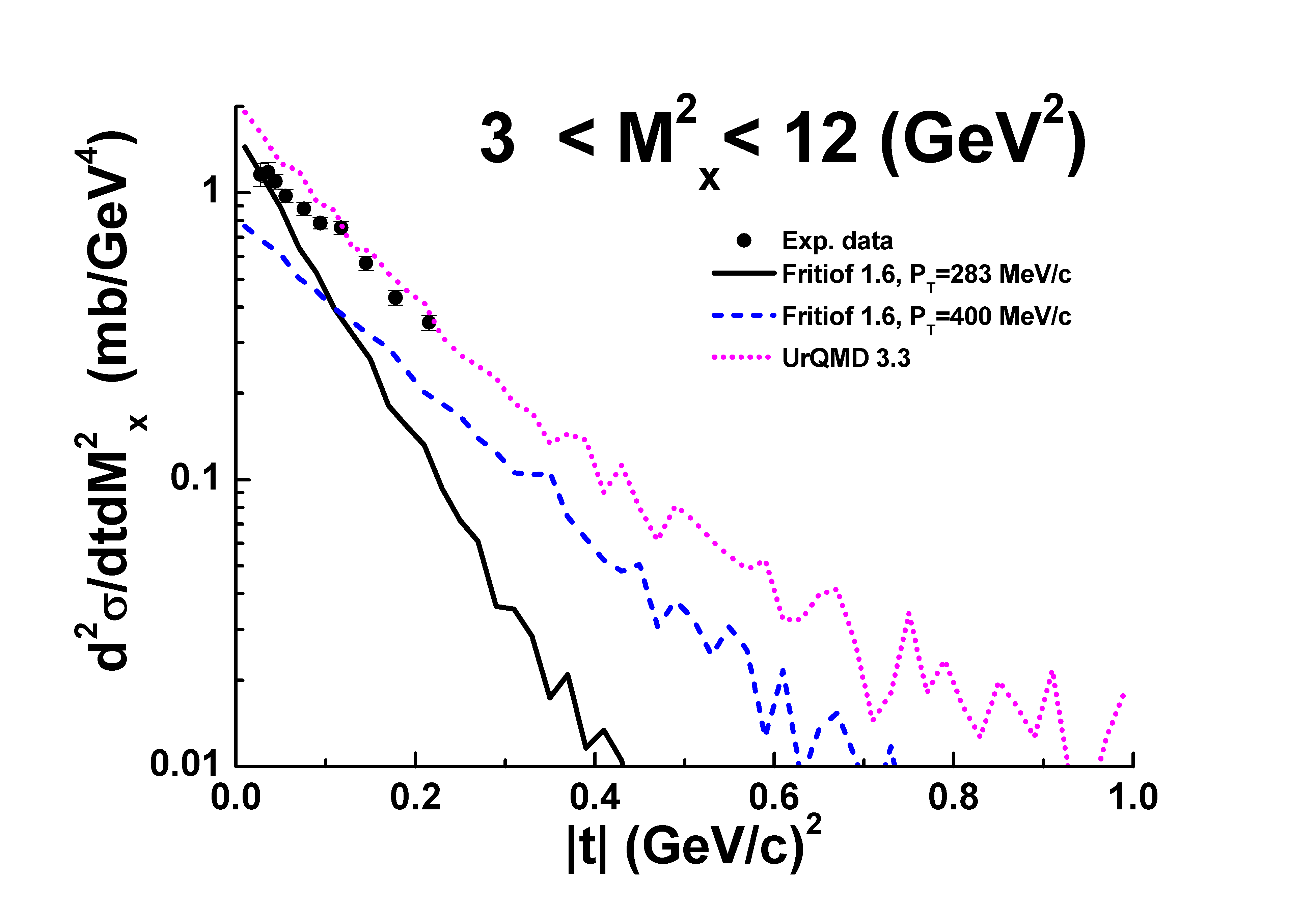}
        \includegraphics[width=80mm,height=35mm,clip]{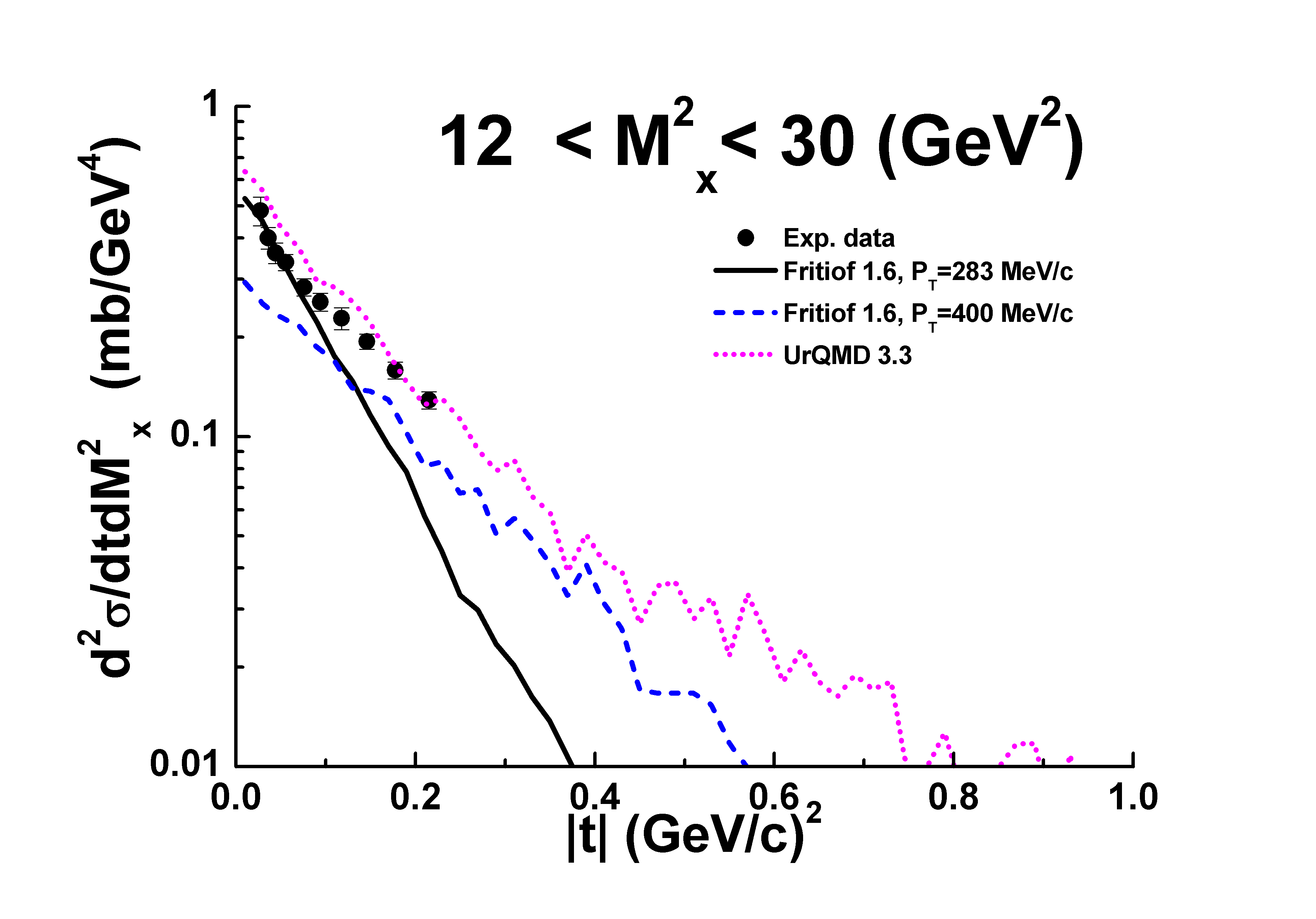}\includegraphics[width=80mm,height=35mm,clip]{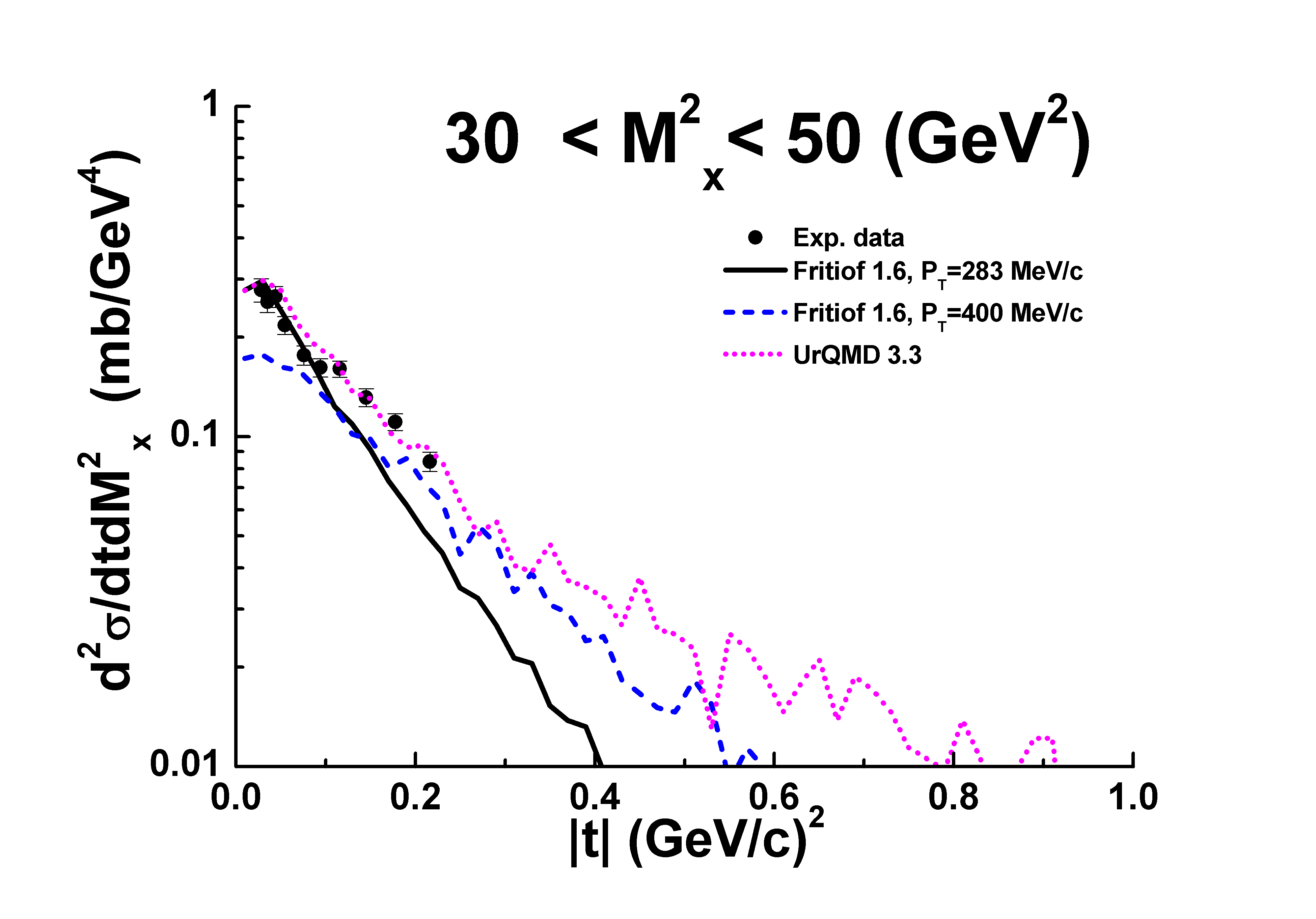}
        \caption{$|t|$ dependence of the cross sections of the reaction $p+p\rightarrow p+X$
                 for 4 intervals in $M^2_x$ at $\sqrt{s}=$23.77 GeV. Points are experimental
                 data of paper \protect{\cite{Scham}} multiplied by 2 for an accounting of
                 the target diffraction. Lines are model calculations.}
        \label{Tdep}
    \end{center}
\end{figure}

It should be noted that $P_T$ distribution in the considered reaction is not determined by
the above given parameter (CTParam(31)=1.6) of the UrQMD model. A final distribution is simulated
in the file angdis.f. Thus, the changes were introduced in it.

Results of the improved UrQMD model are presented in Fig.~\ref{Tdep}.
As seen, the model underestimates a production of systems
with $M_x^2<$2.5 GeV$^2$. At higher mass, the model predictions are in an agreement
with the experimental data.

\begin{figure}[cbth]
    \begin{center}
\includegraphics[width=50mm,height=50mm,clip]{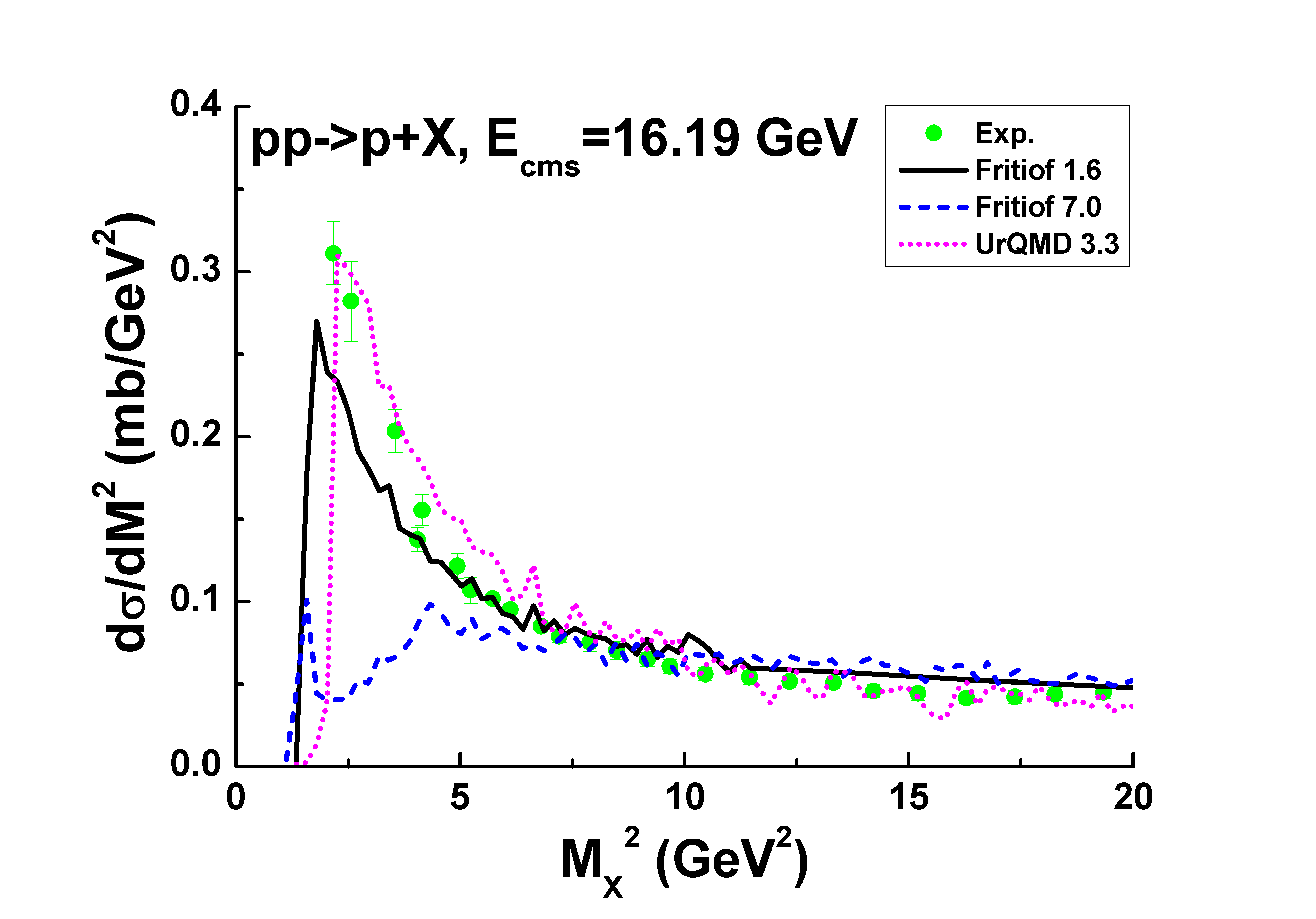}\includegraphics[width=50mm,height=50mm,clip]{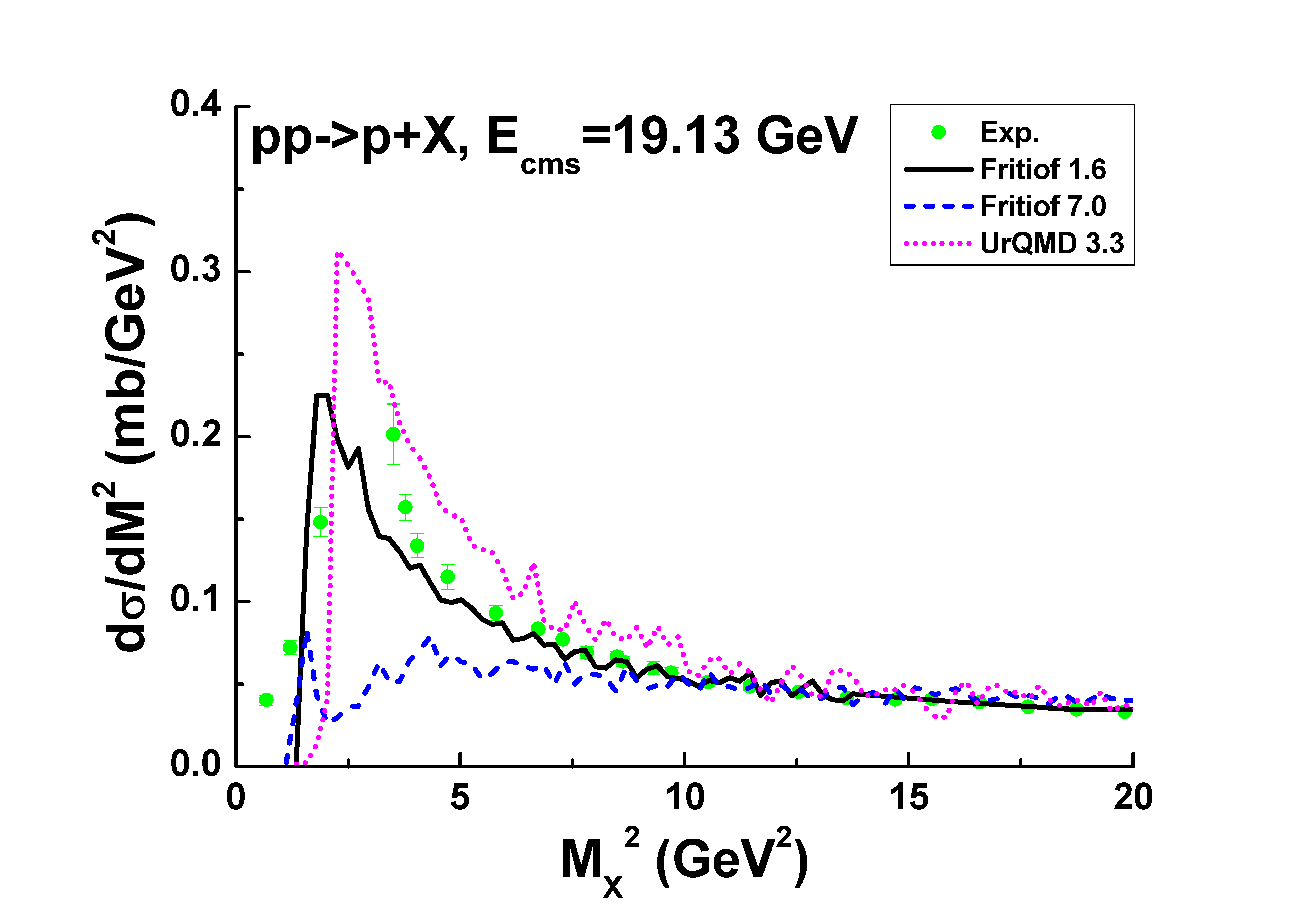}\includegraphics[width=50mm,height=50mm,clip]{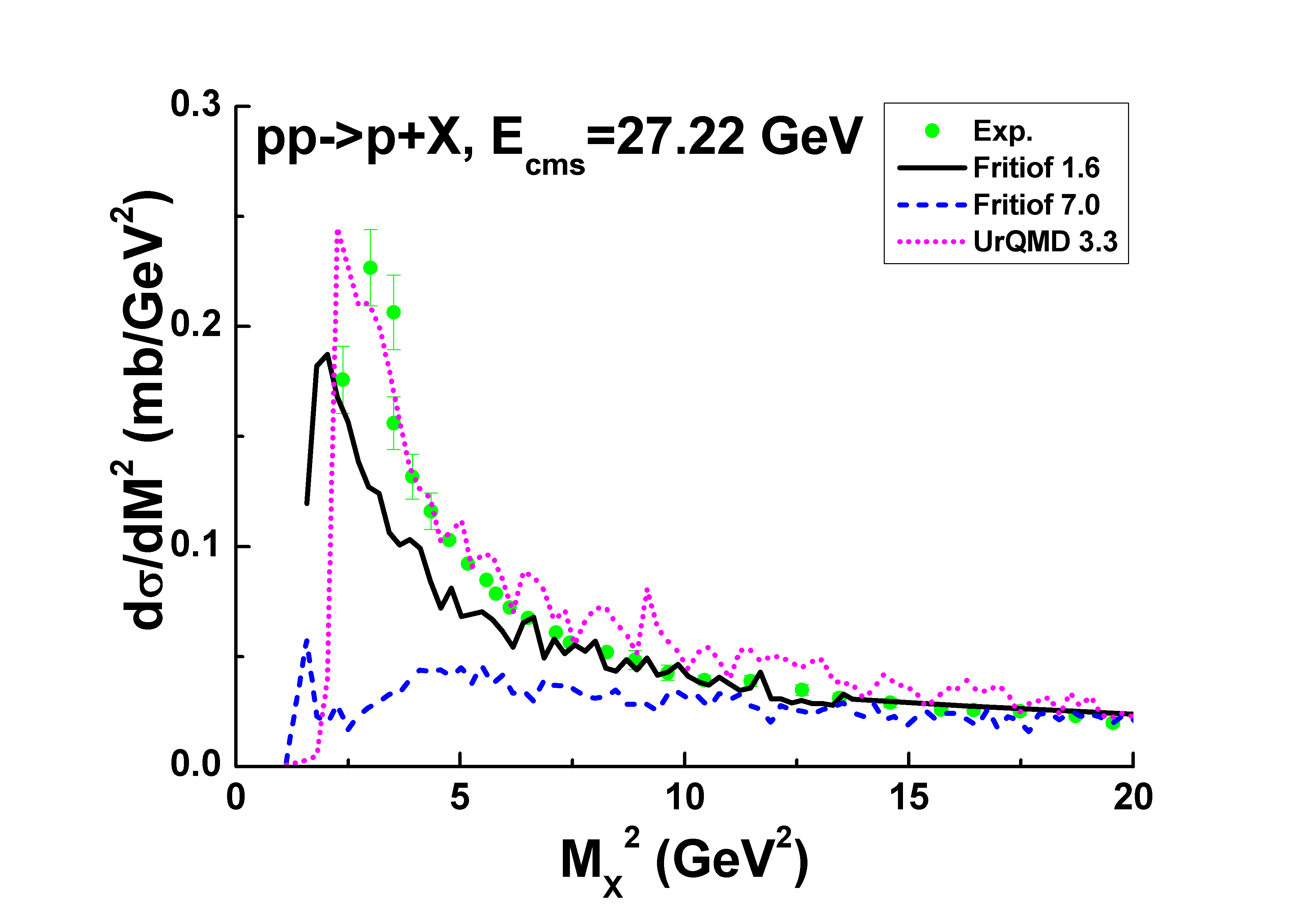}
        \caption{Squared mass distributions in the reactions $p+p\rightarrow p+X$.
           Points are experimental data \protect{\cite{Scham}}. Lines are model calculations.}
        \label{Mx2Distr}
    \end{center}
\end{figure}

Assuming that a correct reproduction of the $t$-dependence of the cross sections is not very
important we can compare the model calculations with other experimental data.
In Fig.~\ref{Mx2Distr} the experimental data of the paper \cite{Scham} on squared mass
distributions in the reactions $p+p\rightarrow p+X$ at various energies integrated over
$|t|$ region 0.024 -- 0.235 (GeV/c)$^2$ are presented in a comparison with the calculations.

\vspace{-5mm}
As seen, the Fritiof 1.6 model does not describe the data in low mass region. The Fritiof
7.0 model gives strange predictions in the region. The UrQMD model reproduces the data
reasonably well except the data at $\sqrt{s}=19.13$ GeV. Probably, more realistic
description of the $t$-dependence is needed. The models have close predictions in high
mass region.

\begin{figure}[cbth]
    \begin{center}
        \includegraphics[width=80mm,height=70mm,clip]{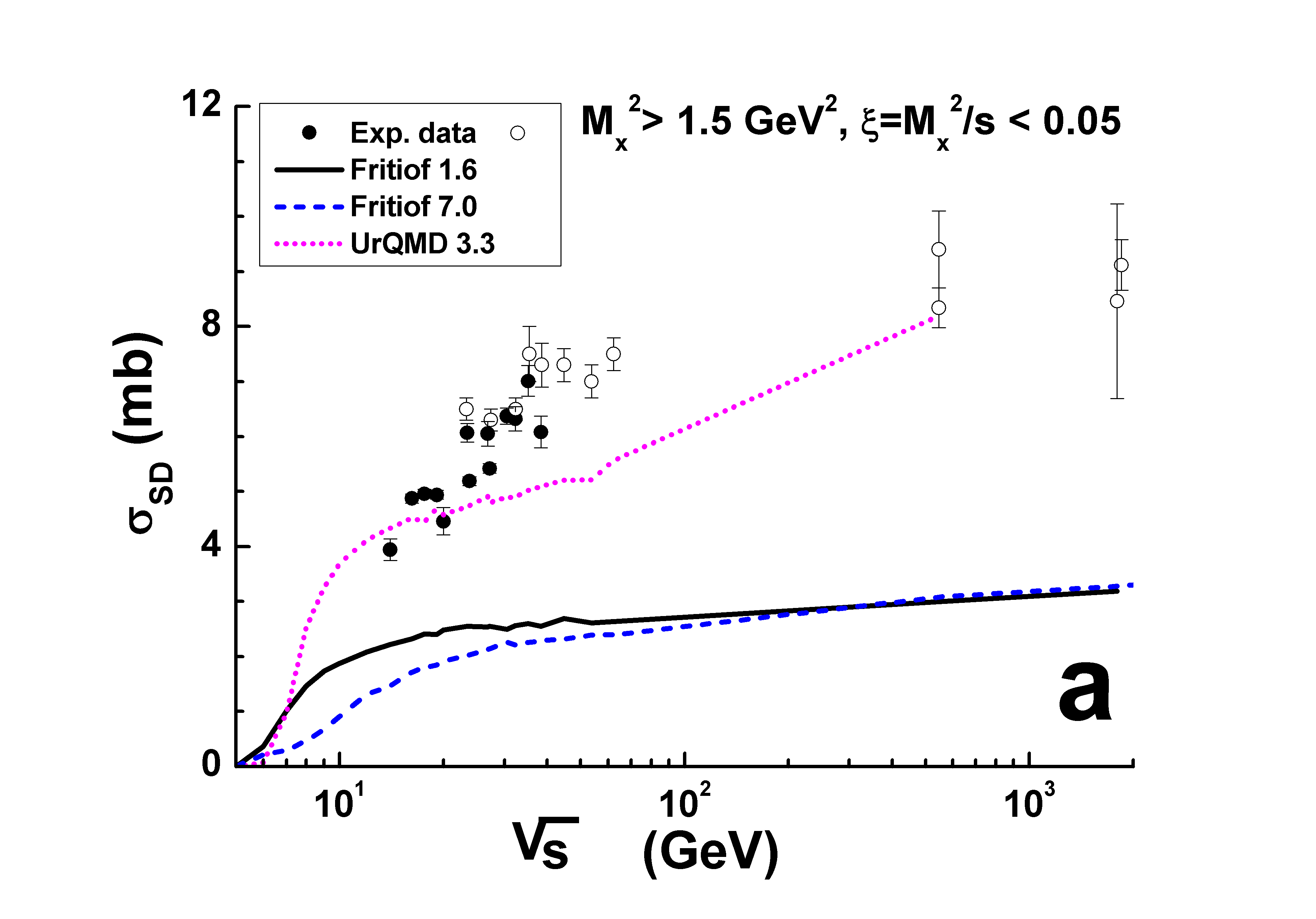}\includegraphics[width=80mm,height=70mm,clip]{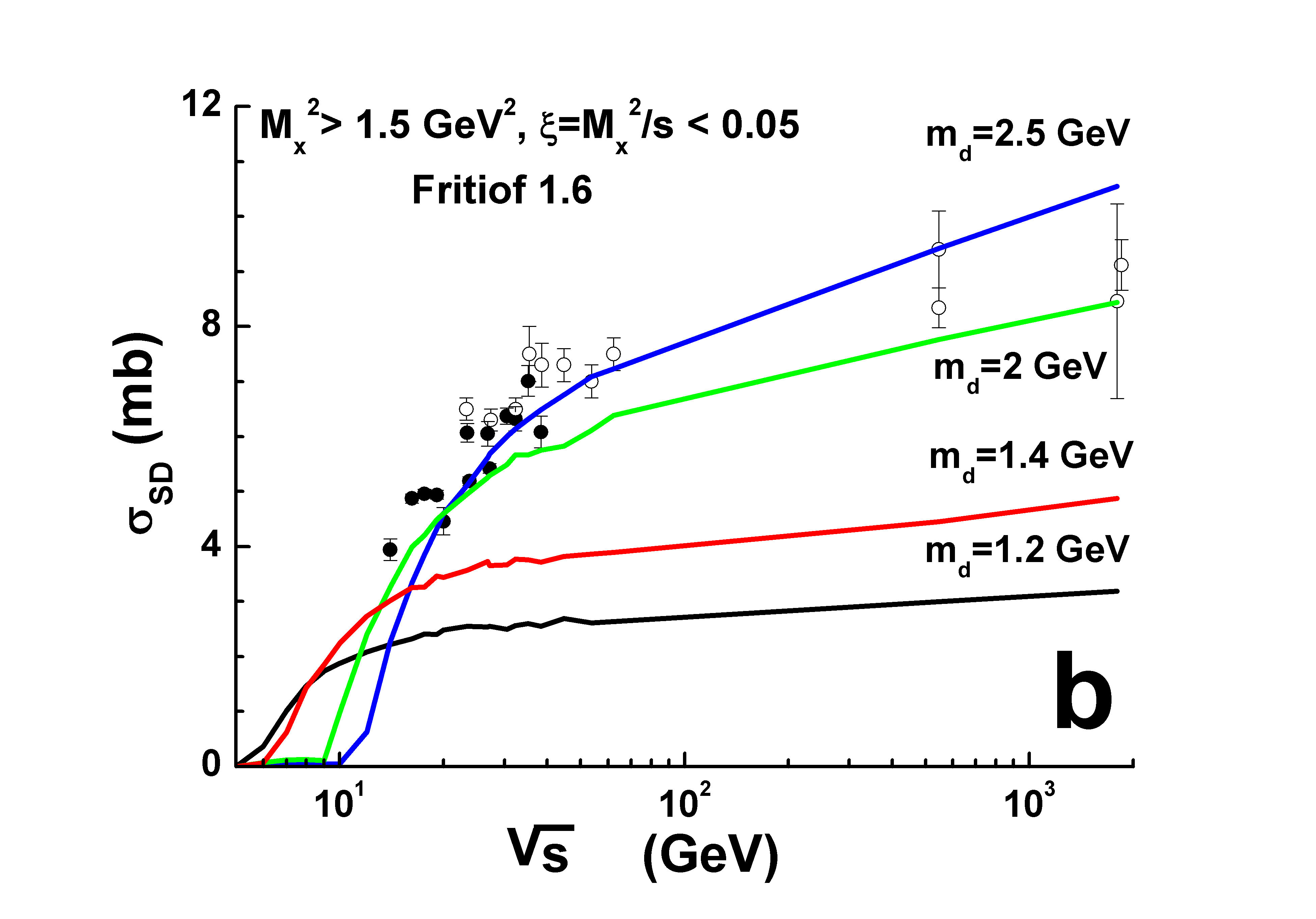}
        \caption{Single diffraction dissociation cross section in $pp$ and $\bar pp$
                 interactions. Points are experimental data presented in
                 \protect{\cite{Goulianos}}. Lines are model calculations.}
        \label{Xdiffr}
    \end{center}
\end{figure}

Calculations of the diffraction dissociation cross sections in $pp$-interactions
integrated over $|t|$ are presented in Fig.~\ref{Xdiffr}a in a comparison with
experimental data from paper \cite{Goulianos}.
As seen, Fritiof 1.6 and Fritiof 7.0 essentially underestimate the cross sections
at high energies. The UrQMD model gives reasonable predictions at
$\sqrt{s}\sim $ 15--25 GeV. Above the region, the model underestimates the data.

Of course, there is a simple possibility to change the Fritiof 1.6 and Fritiof 7.0
predictions increasing $m_d$ (see Fig.~\ref{Xdiffr}b). This changes also the behaviour
of the cross sections at low energies. However the particle production in the central
region of $pp$-interactions has a weak dependence on the ratio between diffractive and
non-difractive cross sections. The most important factor here is a particle production
in non-diffractive interactions.

\section{$p{\rm C}$ interactions}
\begin{figure}[cbth]
\includegraphics[width=53mm,height=50mm,clip]{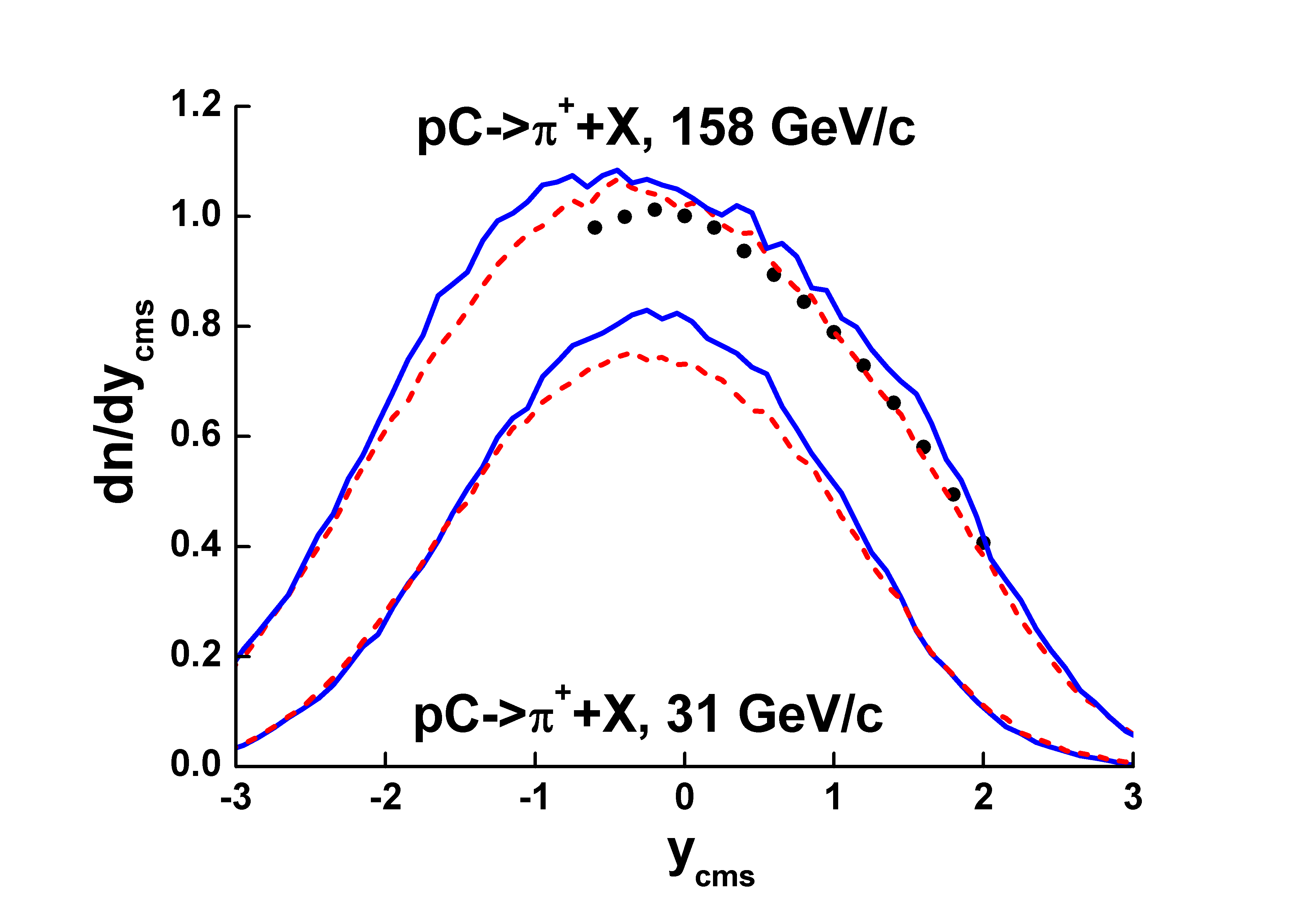}\includegraphics[width=53mm,height=50mm,clip]{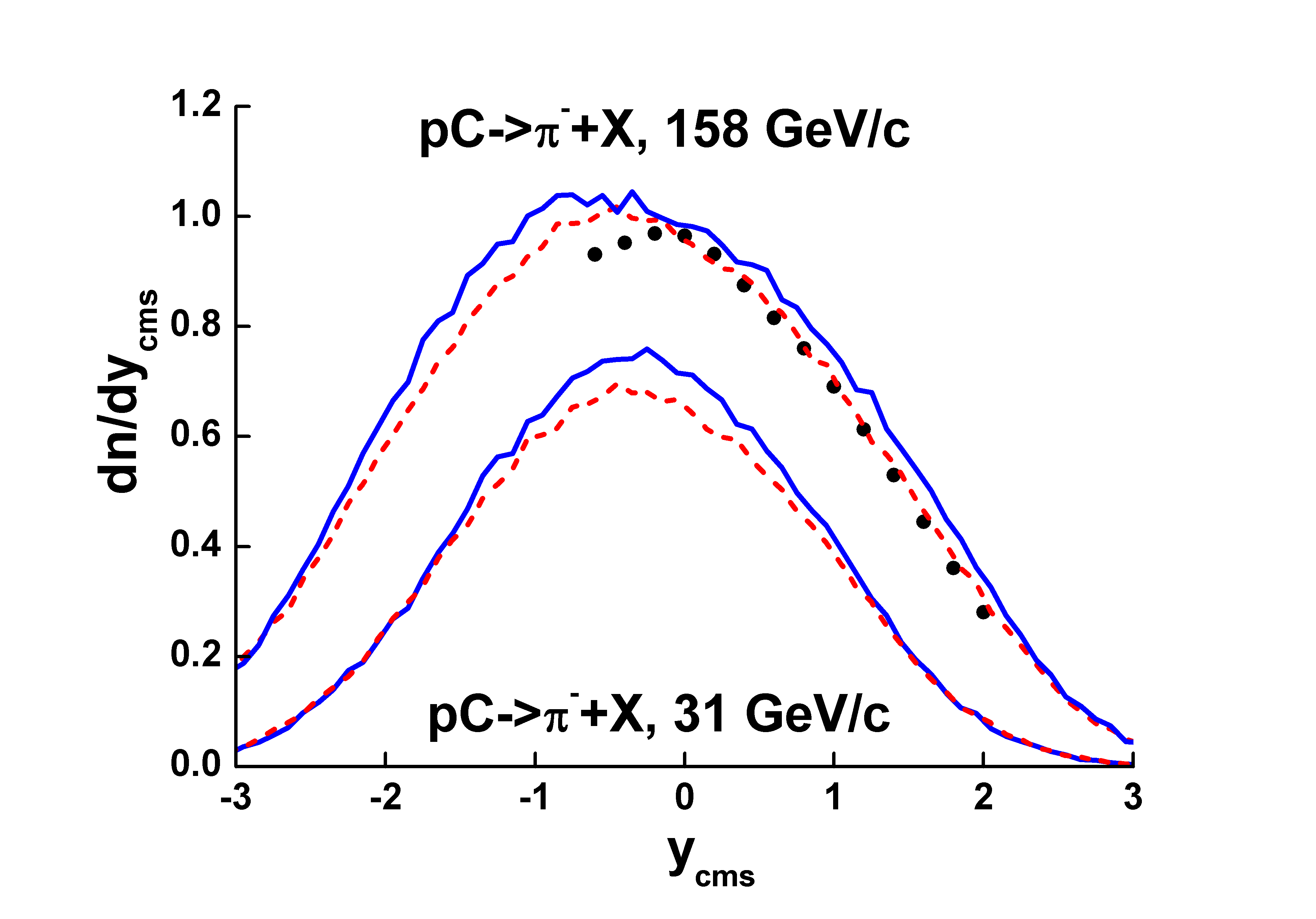}\includegraphics[width=53mm,height=50mm,clip]{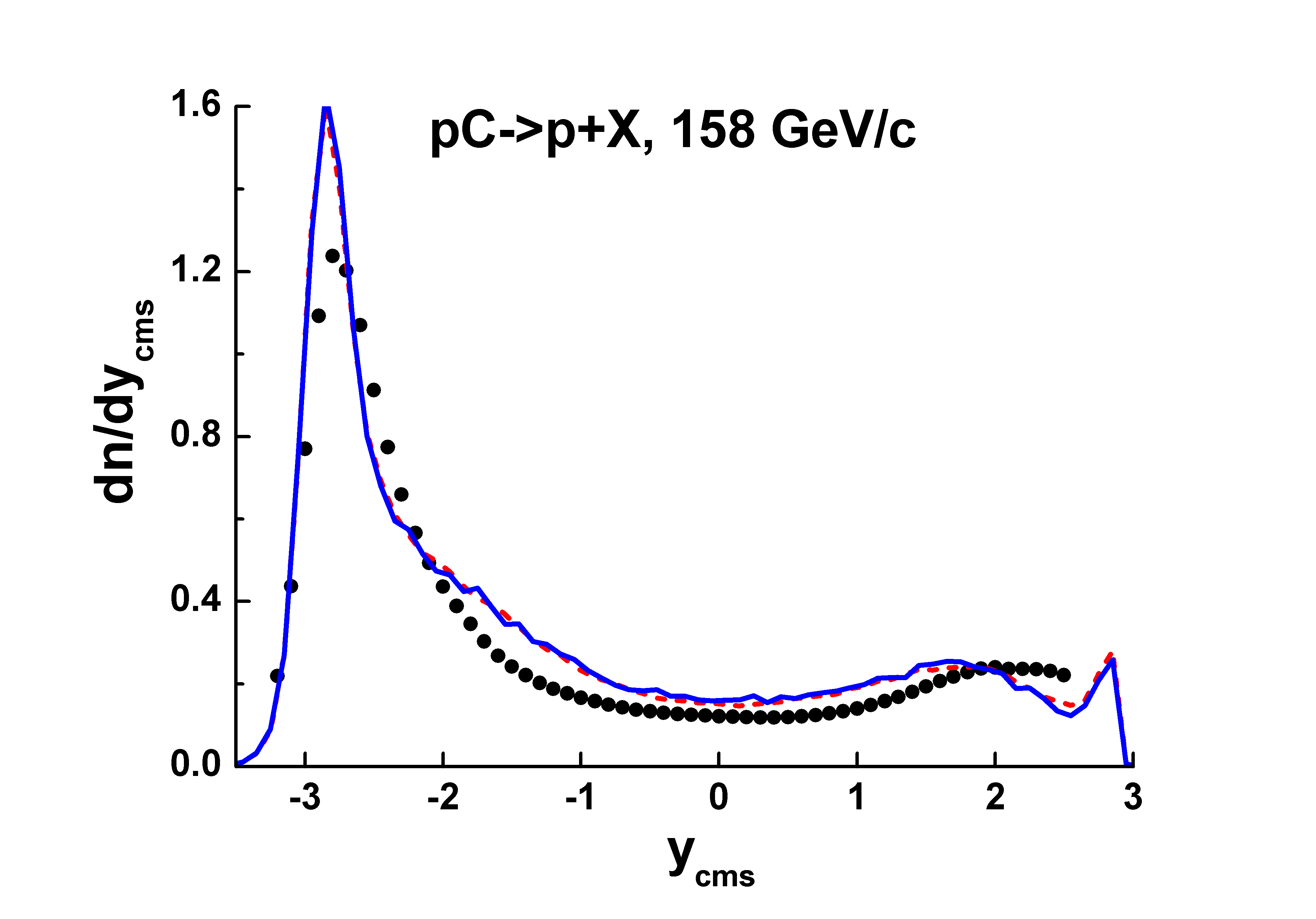}
\caption{Rapidity distributions of mesons and protons in $p{\rm C}$ interactions.
Points are experimental data \protect{\cite{NA49pp2pi,NA49pp2p}} at $P_{lab}=$ 158 GeV/c.
Lines are UrQMD model calculations: solid lines are the calculations with tuned probabilities
of the Fritiof processes, dashed lines are the standard model calculations.
Upper and lower curves are calculations results at $P_{lab}=$ 158 and 31 GeV/c, respectively.}
\label{Ur_pC}
\end{figure}

The most simple situation takes place in the UrQMD model with a description of the $p{\rm C}$
data at $P_{lab}=$ 158 GeV/c. Because the standard and tuned variants of the model give close
results, and a number of intra-nuclear collision in $p{\rm C}$ interactions is not large
($\sim 1.4$), one cannot expect a large difference for $p{\rm C}$ collisions. According to
Fig.~\ref{Ur_pC} it is so.

The difference becomes larger at lower energies, but it is lower than in $pp$-interactions.
Thus one can suppose that the tuning is not important for nucleus-nucleus interactions.

According to the figure, the bad situation with the description of the proton spectrum in
$pp$ interactions is saved for $p{\rm C}$ interactions also.

\begin{figure}[cbth]
\includegraphics[width=53mm,height=50mm,clip]{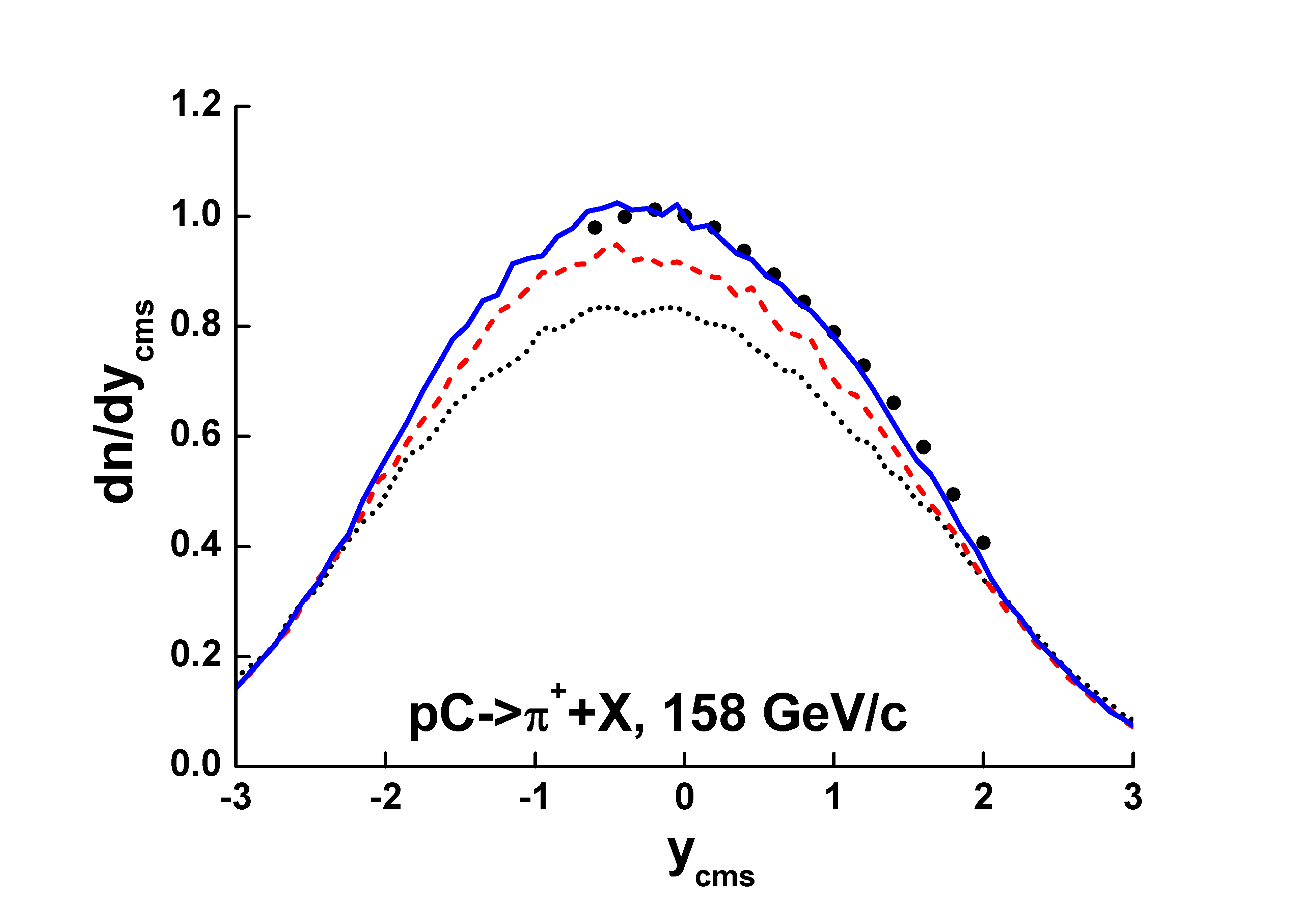}\includegraphics[width=53mm,height=50mm,clip]{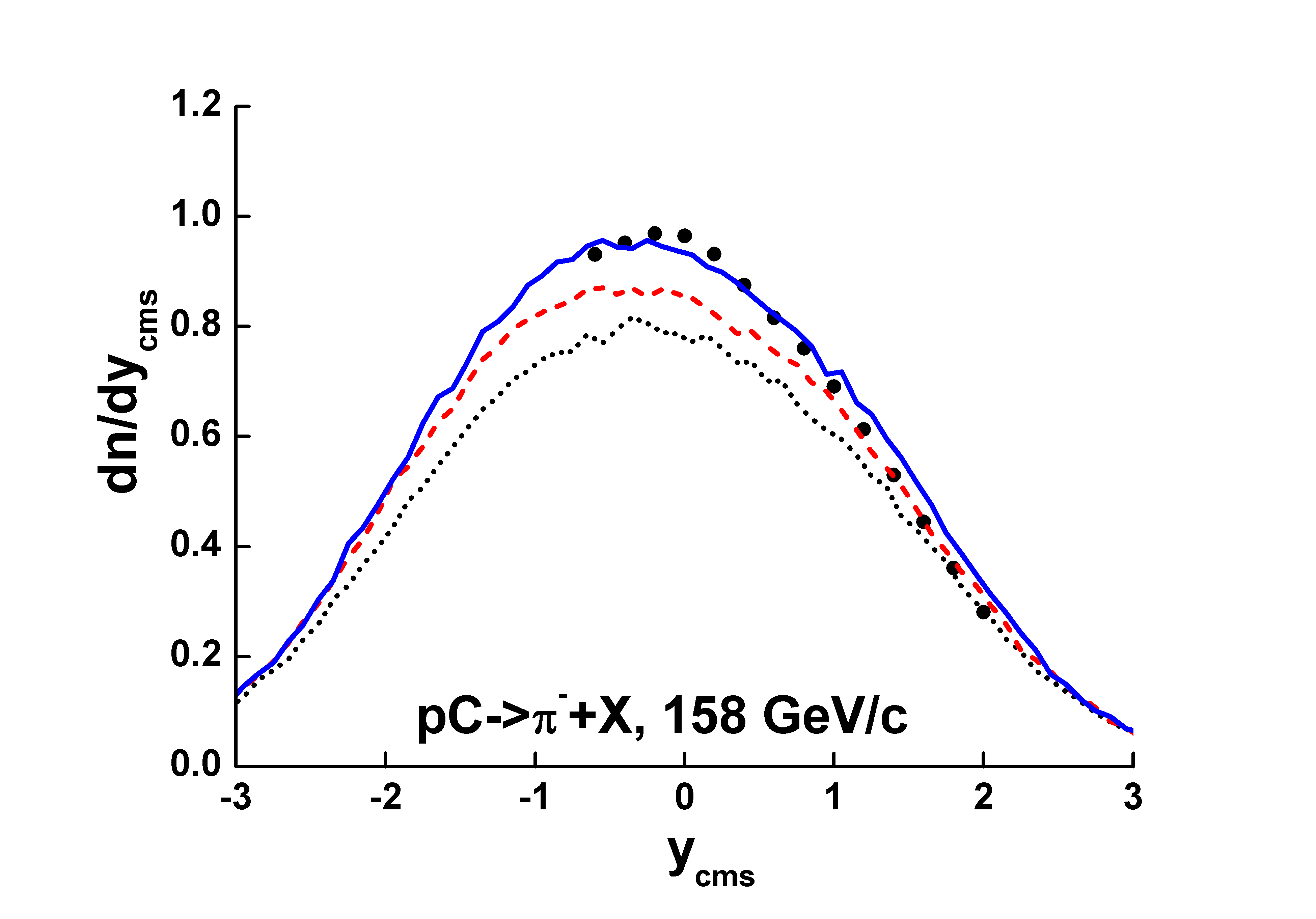}\includegraphics[width=53mm,height=50mm,clip]{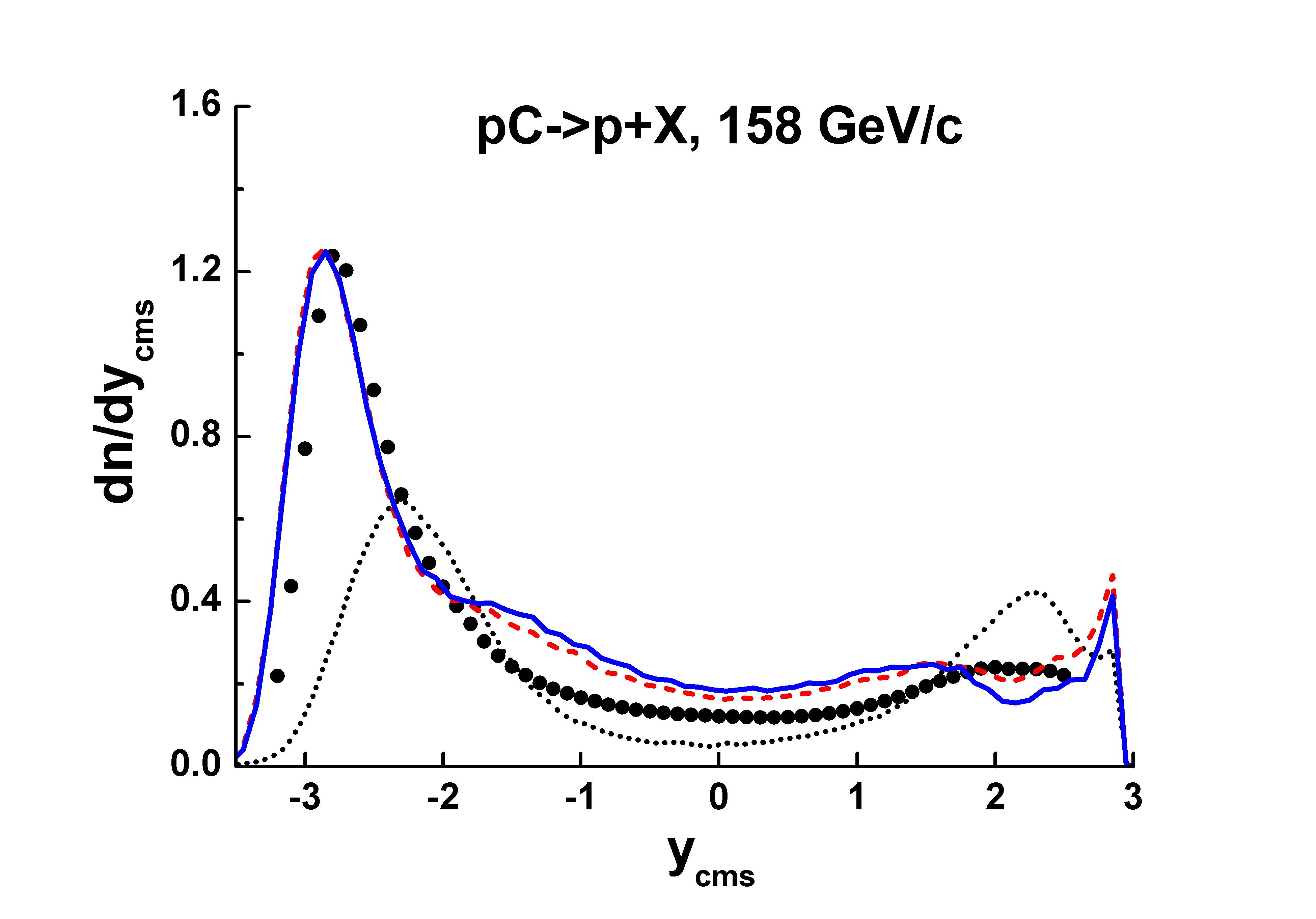}
\caption{Rapidity distributions of mesons and protons in $p{\rm C}$ interactions.
Points are experimental data \protect{\cite{NA49pp2pi,NA49pp2p}} at $P_{lab}=$ 158 GeV/c.
Lines are Fritiof 1.6 model calculations: dashed and solid lines are the calculations with
and without the de-excitations, dotted lines are the standard model calculations.}
\label{Fr_pC}
\end{figure}

Results of the tuned Fritiof 1.6 model are presented in Fig.~\ref{Fr_pC}. There are
two variants of the model for hadron-nucleus and nucleus-nucleus interactions -- with
and without de-excitations of the created objects during intra-nuclear collisions.
The variant without the de-excitations assumes that a mass of an object can only increase
during the collisions, due to this a multiplicity of produced particles is also increased.
As seen, the variant without the de-excitations allows to decribe the data.

As it was said above, the original version of the model did not consider the secondary
particle cascading into nuclei. For a simulation of it, a reggeon theory inspired model
\cite{RTIM} of nuclear destruction (RTIM) was attracted and implemented in the tuned variants.
As seen, this allows to describe the proton spectrum in the target fragmentation region.
A good description of the spectrum in the central region requires a good reproduction of
the spectrum in $pp$-interactions.

\section*{Conclusion}
\begin{enumerate}
\item Simulations of the low mass diffraction dissociation is not correct enough in all
Fritiof-based models.

\item It would be well to implement a simulation of the binary reaction in all Fritif-based
model, and improve their cross in the UrQMD model.

\item It is no doubts that due to a fine tuning of the model's parameters a successful
description of meson production experimental data can be reached.

\item The problem with the proton spectra is left. Maybe, it will be needed to account
for the Gribov inelastic screening in hadron-nucleus interactions (see references in
\cite{GIS1,GIS2}) for a problem solution.
\end{enumerate}

\end{document}